\documentclass[prl,twocolumn,nofootinbib,superscriptaddress]{revtex4-1}

\usepackage{xcolor}
\usepackage{hyperref}
\hypersetup{
	colorlinks,
	linkcolor={red!90!black},
	citecolor={black!10!blue},
	urlcolor={blue!80!black}
}

\usepackage{amssymb}
\usepackage{bm,amsmath}
\usepackage{graphicx}
\usepackage{array}

\newcommand{\f}{\frac}

\begin{document}

\title{Non-trivial activity dependence of static length scale and critical tests of active random first-order transition theory}

\author{Kallol Paul}
\affiliation{TIFR Centre for Interdisciplinary Sciences, Tata Institute of Fundamental Research, Hyderabad - 500046, India}

\author{Saroj Kumar Nandi}
\affiliation{TIFR Centre for Interdisciplinary Sciences, Tata Institute of Fundamental Research, Hyderabad - 500046, India}

\author{Smarajit Karmakar}
\affiliation{TIFR Centre for Interdisciplinary Sciences, Tata Institute of Fundamental Research, Hyderabad - 500046, India}

\begin{abstract}
Effects of activity on glassy dynamics are fundamental in several biological processes. Active glasses extend the scope of the equilibrium problem and provide new control parameters to probe different theoretical aspects. In the theory of glassy dynamics, different length scales play pivotal roles. Here, for the first time, we present results for the static length scale, $\xi_S$, in an active glass via large-scale molecular dynamics simulations for model active glasses in three spatial dimensions. We show that although the relaxation dynamics are equilibrium-like, activity has non-trivial effects on $\xi_S$. $\xi_S$ plays the central role in the random first-order transition (RFOT) theory. Thus, our work provides critical tests for the active RFOT theory, a phenomenological extension of its equilibrium counterpart. We find that the two exponents, $\theta$ and $\psi$, within the theory, become activity-dependent, exposing the non-trivial effects of activity on $\xi_S$. However, the combination of $\theta$ and $\psi$, which controls the relaxation dynamics, remains nearly independent of activity leading to the effectively equilibrium-like behavior. Interestingly, $\xi_S$ shows higher growth in an active glass; this should help better comparison of theories with simulations and experiments.
 \end{abstract}

\maketitle

Glassy dynamics is a hallmark of nearly any liquid; when supercooled below their melting temperature \cite{giulioreview}, for a modest decrease in temperature, $T$, or increase in density, a glassy system shows an impressive growth -- from picoseconds to hours -- of relaxation time, $\tau$. The microscopic mechanism leading to this tremendous slow down in the supercooled regime continues to be debated, leading to various theories for the phenomenon \cite{giulioreview,lubchenko2007,das2004,dyre2009,giuliobook,tarjus2005,smarajitAnnual}. Most of these theories posit a transition at a lower $T$ that is inaccessible to simulations or experiments, and thus, definitive tests of different theories are complex. For example, random first-order transition (RFOT) theory predicts a thermodynamic transition at $T_K$ where the relaxation time diverges. From this aspect, active glasses, where the systems consist of particles with a self-propulsion force, $f_0$, and a persistence time, $\tau_p$, of their motion \cite{sriramreview,sriramrmp}, are significant since they extend the scope and extent of the problem. Activity can provide an additional probe for glass transition. Furthermore, they are fundamental in several biological processes, such as wound healing \cite{poujade2007,olivier2014}, collective migration in a cell monolayer \cite{angelini2011,park2015,malinverno2017,garcia2015}, intracellular dynamics \cite{parry2014,nishizawa2017,zhou2009}, tumor progression \cite{kakkada2018,streitberger2020}. Consequently, various models of such systems have been simulated \cite{berthier2014,kallol2021,mandal2016,flenner2016}, and theories of equilibrium glasses, such as the mode-coupling theory \cite{activemct,berthier2013,feng2017,szamel2016} and the random first-order transition (RFOT) theory \cite{activerfot}, have been extended for such systems.

How does the presence of nonequilibrium activity in a system affect the aspects of glassy behavior? Intensive research of the last decade seems to suggest that apart from some quantitative changes, such as the shift in the glass transition temperature, $T_g$ \cite{ni2013,berthier2013,activemct}, the qualitative aspects of glassy dynamics in an active and passive system remain similar. This similarity implies that active glasses can be described much like an equilibrium system, via an effective temperature, $T_{eff}$. Such a similarity, however, is somewhat surprising and disappointing at the same time. Activity is known to show nontrivial effects, such as flocking \cite{vicsek1995}, giant number fluctuation \cite{vijay2007,giavazzi2017}, and long-range velocity correlation \cite{caprini2020,szamel2016,berthier2019} in a dilute system. Some of these effects survive in the glass regime; however, it would be surprising if the activity had no nontrivial effects on the glassy properties. In addition, it is disappointing as activity could provide a worthwhile probe to understand the nature of equilibrium glasses around $T_K$. Specifically, activity allows access to the system properties close to $T_K$, much like random pinning \cite{skPhysica2012,chiara2012,sourishSciReport,softMatter2017,OzawaPNAS} that enables one to study the system properties close to the ideal glass transition in equilibrium systems. Now, suppose activity merely shifts the transition point. In that case, the problem remains similar to an equilibrium system: one has to go to the renormalized transition point to study the properties of the ideal glass. In a recent paper we showed, that although the relaxation dynamics are qualitatively similar to an equilibrium system, activity has nontrivial effects on the dynamic heterogeneity (DH) length scale, $\xi_D$ \cite{kallol2021}. Our extended mode-coupling theory (MCT) of active glasses agrees well with these behaviors. Does then activity also have nontrivial effects on the other length scales, specifically, the static length scale, $\xi_S$, that controls the glassy dynamics within RFOT? If yes, how can we understand the origin of the effectively equilibrium-like behavior of the relaxation dynamics?

In this work, we carry out large-scale molecular dynamics simulations of a well-known, three-dimensional model of active glass, the Kob-Anderson binary mixture with particle ratio $80:20$, and take $10\%$ of the particles as active \cite{kallol2021}. The motivation to take a fraction of the particles as active comes from biological systems such as the actomyosin cortex or cellular cytoplasm, where a fraction of proteins (such as the myosin molecules) are active \cite{nishizawa2017,parry2014,jacques2015}. The qualitative results are independent of this fraction as long as it is non-zero \cite{kallol2021}. In our simulations, we have implemented activity characterized as model-1 in the language of Ref. \cite{activerfot}. We have investigated the role of activity by varying $f_0$ in our simulation, keeping $\tau_p=1$ fixed. From the perspective of $f_0$, both models, discussed in Ref. \cite{activerfot}, are similar; thus, the results presented here are applicable for both models of activity. We calculated $\xi_S$ via two distinct methods, the finite-size scaling \cite{smarajitPNAS2009} and the block analysis \cite{saurish2017, ACSOmegaReview}, and find they give nearly the same results. Here, we report the first measurements of $\xi_S$ in active systems and show that indeed activity has nontrivial effects on $\xi_S$. 
The work presented here provides critical tests of the active RFOT theory of Ref. \cite{activerfot}. Specifically, we find that the assumption of the exponents being independent of activity is inappropriate. However, their combination that appears within the theory for the relaxation dynamics is independent of activity, consistent with Ref. \cite{activerfot}: this leads to the effectively equilibrium-like behavior of the relaxation dynamics.



RFOT theory is one of the most popular theories of glass transition and has been remarkably successful in describing the glassy aspects in diverse systems. Within RFOT theory \cite{lubchenko2007,kirkpatrick2015}, a glassy system consists of mosaics of different states. Consider a region of length scale $R$. The state within this region is set by two competing effects: the surface energy cost of having different states and the bulk energy. Within RFOT theory, the reconfiguration is entropic in nature, thus $f=Ts_c(\tilde{\Phi},T)$ where$s_c$ is the configurational entropy, $\tilde{\Phi}$ is the effective interatomic interaction, and $T$, the temperature. Thus, the free energy cost can be written as
\begin{equation}
\Delta F=-\Omega_d R^d Ts_c(\tilde{\Phi},T)+S_d R^{\theta}\bar{\gamma},
\end{equation}
where $\Omega_d$ and $S_d$ are the volume and surface of a hypersphere in spatial dimension $d$. Minimizing $\Delta F$, we obtain the typical length scale of the mosaics as $\xi_S\sim[c_d\bar{\gamma}/Ts_c(\tilde{\Phi},T)]^{1/(d-\theta)}$, where $c_d=\theta S_d/d\Omega_d$. The energy cost, $\bar{\gamma}$ is proportional to $T$: $\bar{\gamma}=\Xi(f_0,\tau_p)T$, where the activity dependence is included in $\Xi(f_0,\tau_p)$. Vanishing of $s_c$ primarily controls the glassy dynamics within RFOT theory; therefore, we ignore the activity dependence of $\Xi$ and treat it as a constant for simplicity. It was shown in Ref. \cite{activerfot} that $s_c(\Phi,T)$ of an active fluid in the small activity regime can be obtained by expanding the contribution of activity around that for the passive system. Thus $s_c(\tilde{\Phi},T)\sim s_c(\Phi,T)+\kappa \delta\Phi$, where $\Phi$ represents the interaction potential for the passive system, $\delta \Phi$ is the potential coming from activity, and $\kappa$ is a constant. For the passive system, close to the Kauzmann temperature, $T_K$, we have $s_c(\Phi,T)\sim \Delta C_p(T-T_K)/T_K$, where $\Delta C_p$ is the difference of specific heats between the crystalline phase to that of the glassy phase at the same $T$. Then, $\xi_S$ becomes
\begin{equation}
\xi_S\simeq \left( \f{D}{T-T_K+\frac{\kappa T_K\delta\Phi}{\Delta C_p}} \right)^{1/(d-\theta)},
\end{equation}
where $D=c_d\Xi T_K/\Delta C_p$ is a constant. Within the mean-field approximation, Ref. \cite{activerfot} obtained $\delta\Phi$ considering a single active particle trapped in a potential created by the surrounding particles. Within the harmonic approximation, it can be analytically shown that $\delta\Phi=Hf_0^2$, where $H$ is a constant \cite{activerfot}. Therefore, writing $K=H\kappa T_K/\Delta C_p$, we obtain
\begin{equation}\label{xieq}
\xi_S\simeq \left( \f{D}{T-T_K+Kf_0^2} \right)^{1/(d-\theta)}.
\end{equation}

\begin{figure}
	\includegraphics[width=8.6cm]{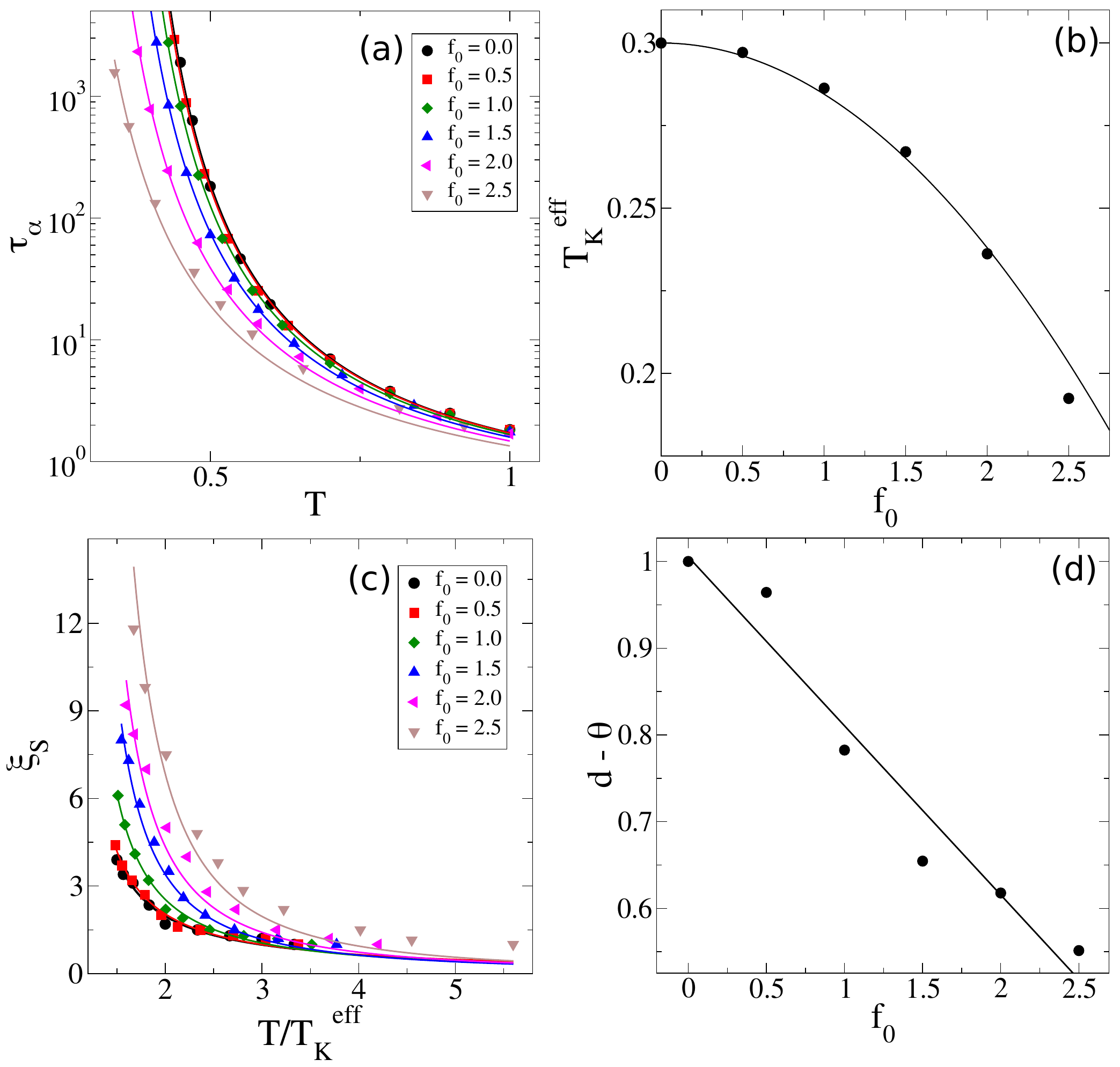}
	\caption{(a) Relaxation time, $\tau$, as a function of $T$ for different $f_0$. Symbols are simulation data and lines are fits with the active RFOT theory, Eq. (\ref{taueq}) with $\psi/(d-\theta)=1$, with $\tau_0=0.263$, $E=1.319$, $T_K=0.3$, and $K=0.0155$. (b) $T_K$ at different $f_0$, symbols represented the values obtained from fitting simulation data with the VFT form, and the line is the active RFOT prediction. (c) Static length scale, $\xi_S$, as a function of $T/T_K$. Symbols are simulation data, and lines represent fits with Eq. (\ref{xieq}). The exponents, $(d-\theta)$ at different $f_0$ are shown in (d) The line is $f(x)=1-0.2 x$.}
	\label{figure1}
\end{figure}

RFOT theory describes relaxation dynamics as the relaxation of individual mosaics. The energy barrier $\Delta$ associated with a region of length $\xi_S$ is $\Delta\sim \Delta_0 \xi_S^\psi$, where $\Delta_0$ is an energy scale, and $\psi$ is another exponent. The precise $T$-dependence of $\Delta_0$ depends on models, following Wolynes {\it et al}, we take it as $\Delta_0=k_B T$ \cite{lubchenko2007,giuliobook} with $k_B$ being the Boltzmann constant. Considering a barrier crossing scenario, we then get the relaxation time, $\tau$ as $\tau=\tau_0 \exp[\Delta_0\xi_S^\psi/T]$, where we have set $k_B$ to unity and the high $T$ value of $\tau$, denoted as $\tau_0$, is independent of $T$ but may depend on activity \cite{activemct}. Then, $\tau$ becomes
\begin{equation} \label{taueq}
\ln\left(\f{\tau}{\tau_0}\right)=\left( \f{E}{T-T_K+Kf_0^2} \right)^{\psi/(d-\theta)},
\end{equation}
where $E=k^{(d-\theta)/\psi}D$.
Further progress demands knowledge of the exponents $\theta$ and $\psi$. Note that a first-principle calculation of $\psi$ does not exist to date even for an equilibrium system \cite{lubchenko2007,giuliobook}. Reference \cite{activerfot} assumed that they remain the same as in the passive system even in the presence of activity and found excellent agreement with simulations. However, this crucial assumption has not yet been tested. Moreover, in the light of our recent work \cite{kallol2021}, where we showed that the exponents of the DH length scale depend on activity: these results suggest activity dependence of both $\theta$ and $\psi$ as well, what we indeed find. However, we show that the activity-dependence of $\theta$ and $\psi$ are such that they cancel each other in the behavior of $\tau$, Eq. (\ref{taueq}).

We now compare the active RFOT theory with our simulation data. Ref. \cite{activerfot} assumed that the $\psi/(d-\theta)=1$ remains valid even in the presence of activity. This assumption was motivated by the observation that the relaxation dynamics remains equilibrium-like; the resulting theory agrees well with the simulation data. Although it has not yet been tested, let us first consider that this assumption holds. For a particular system, the value of $K$ should be the same. Analysis presented in Ref. \cite{kallol2021} show that for this particular system, $K=0.0155$. It is also known that $T_K=0.3$ for the passive system. Then, we obtain $\ln(\tau/\tau_0)=E/(T-T_K^{eff})$, where $T_K^{eff}=T_K-Kf_0^2$ is the effective Kauzmann temperature (Eq. \ref{taueq}). Figure \ref{figure1}(a) shows the fits of the simulation data for $\tau$ with this form. $\tau_0$ and $E$ are obtained by fitting the data for the passive system and are kept the same for the active systems. We obtain $T_K^{eff}$ as a fitting parameter. Figure \ref{figure1}(b) shows the comparison of $T_K^{eff}$, obtained via fitting the simulation data, with that of the theory. Note that the line is not a fit.

We now discuss the main result of this paper. It is the behavior of the individual exponents, $\theta$ and $\psi$, in the presence of activity. Note that the precise values of these parameters, even for the equilibrium systems, remain controversial. For a passive system, one school of thoughts have argued \cite{kirkpatrick1989,lubchenko2007,kirkpatrick2015} that $\theta=\psi=d/2$. However, the other possibilities, such as $\theta\sim (d-1)$ and $\psi\lesssim 1$, also look reasonable when fitted with simulation and experimental data \cite{giuliobook}. For our system, the passive system data seem to be well-described by the latter, and we chose these values for the passive system. Using this, fitting Eq. (\ref{xieq}) with simulation data gives $D$, note that $T_K^{eff}$ is already known. Next, we fit Eq. (\ref{xieq}) with with the simulation data for active systems using $(d-\theta)$ as a fitting parameter and show them in Fig. \ref{figure1}(c). The obtained values of ($d-\theta$) are shown in Fig. \ref{figure1}(d); $\theta$ almost linearly increases with $f_0$. One interesting feature is that $\theta$ becomes larger than $(d-1)$ as activity increases. In an equilibrium system, $\theta$ has the upper limit of $(d-1)$. This bound is perhaps not applicable for our system, which is out of equilibrium. What does this progressive increase in $\theta$ with $f_0$ imply? The exponent $\theta$ characterizes the surface of mosaics. As activity increases, $\theta$ tends closer to $d$, implying that the surface becomes quite irregular and assumes a fractal nature.

\begin{figure}
\includegraphics[width=8.6cm]{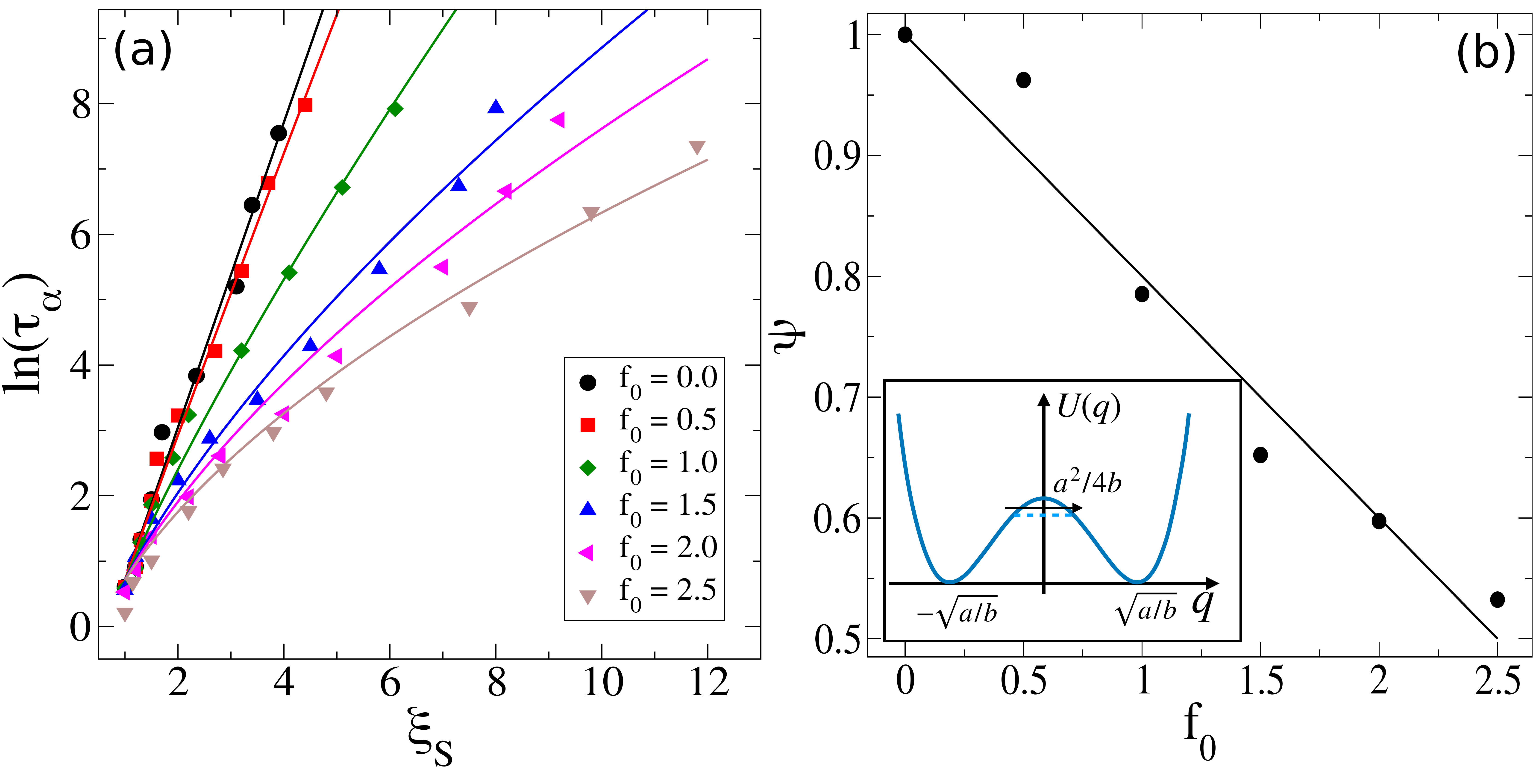}
\caption{(a) Fit of $\ln\tau=c_1+c_2\xi_S^\psi$ for the passive system, using $\psi=1$ gives the values of $c_1=-1.624$ and $c_2=2.335$. Once these parameters are determined, we use $\psi$ as a parameter and fit the data for the active system. The symbols are our simulation data, and the lines are the fits. (b) The symbols show the values of $\psi$ as a function of $f_0$ obtained via fitting. The line is $f(x)=1-0.2x$. {\bf Inset:} Barrier crossing scenario for an active particle (see text for details). }
\label{psixitau}
\end{figure}

We next look at the exponent $\psi$. From the discussion preceding Eq. (\ref{taueq}), we obtain $\ln\tau =c_1+c_2 \xi_S^\psi$, where $c_1$ and $c_2$ are two constants. Figure \ref{psixitau}(a) shows the fits of this function with the simulation data. We have used $\psi=1.0$ for the passive system, although note that a slight variation (0.8 to 1.0) in this value does not affect the fit noticeably. Fits with the passive system data provides the values of $c_1\simeq-1.624$ and $c_2\simeq 2.335$. We then use only $\psi$ as the fitting parameter in the rest of the fits for the active system data. Figure \ref{psixitau}(b) shows the values of $\psi$, obtained from the fits. Note the remarkably similar trend of $\psi$ with that of $(d-\theta)$ in Fig. \ref{figure1}(d).

To understand the effective barrier and modification of $\psi$, let us consider a barrier crossing mechanism in one dimension. For concreteness, consider a two-well potential, $U(q)=-(a/2)q^2+(b/4)q^4+a^2/4b$. The potential is characterized by the two minima at $q=\pm\sqrt{a/b}$ and a maxima at $q=0$ with the barrier height $a^2/4b$. Consider the barrier crossing of the system from one minimum to the other in the presence of the active noise, $\zeta$. Since we are interested in the role of active force, ignoring the thermal noise, we write the equation of motion of the particle as
\begin{align}
\gamma \dot{q}(t) &=F(q)+f_0\zeta(t) \\
\tau_p\dot{\zeta}(t) &=-\zeta(t)+\sqrt{\tau_p}\eta(t)
\end{align}
where $F(q)=\partial_qU(q)$, $\gamma$ is the friction force, $\eta(t)$ is a Gaussian white noise with zero mean and $\langle\eta(t)\eta(t')\rangle=\delta(t-t')$. With a slight algebraic manipulation, we can write the above two equations as
\begin{align}
\gamma\tau_p\ddot{q}=-[\gamma-\tau_p\partial_qF(q)]\dot{q}+F(q)+f_0\sqrt{\tau_p}\eta(t).
\end{align}
The above equation essentially describes a barrier crossing scenario with effective friction $\Gamma=\gamma-\tau_p\partial_qF(q)$ and the strength to cross the barrier is provided via active force. $\Gamma$ shows that the effective friction to cross the barrier reduces when $\tau_p$ is appreciably larger compared to the time-scale of thermal noise. However, it also depends on the nature of the barrier, in particular on $\partial_q^2 F(q)$. Thus as the particle approaches the barrier maxima, the second derivative becomes progressively smaller, and so does the effective friction that becomes negative beyond a certain point. Thus the barrier effectively exists till this point. Even in the absence of thermal noise, activity can provide the required fluctuations to cross the barrier. In our case, when we are varying $f_0$ alone, the effective barrier reduction is provided by both $f_0$ and $\xi_S$. Since $\xi_S$ is not very large in a glassy system, and the minimum value is 1, we propose the following form: $1-cf_0(\xi_S-1)\simeq 1-cf_0\ln\xi_S$. In the limit of small activity, we can write this as $\xi_S^{-cf_0}$. Thus the effective barrier height becomes $\xi_S^{\psi-cf_0}$; this is the form that we have used to fit the variation in $\psi$, Fig. \ref{psixitau}(b). Note that this simple argument does not work when $\tau_p$ is very small, and the scenario resembles the thermal case. On the other hand, when $\tau_p$ is very large, such that $\Gamma$ remains almost always negative, other distinct types of phenomena, resembling jammed systems \cite{mandal2020}, are expected to emerge.

\begin{figure}
	\includegraphics[width=8.6cm]{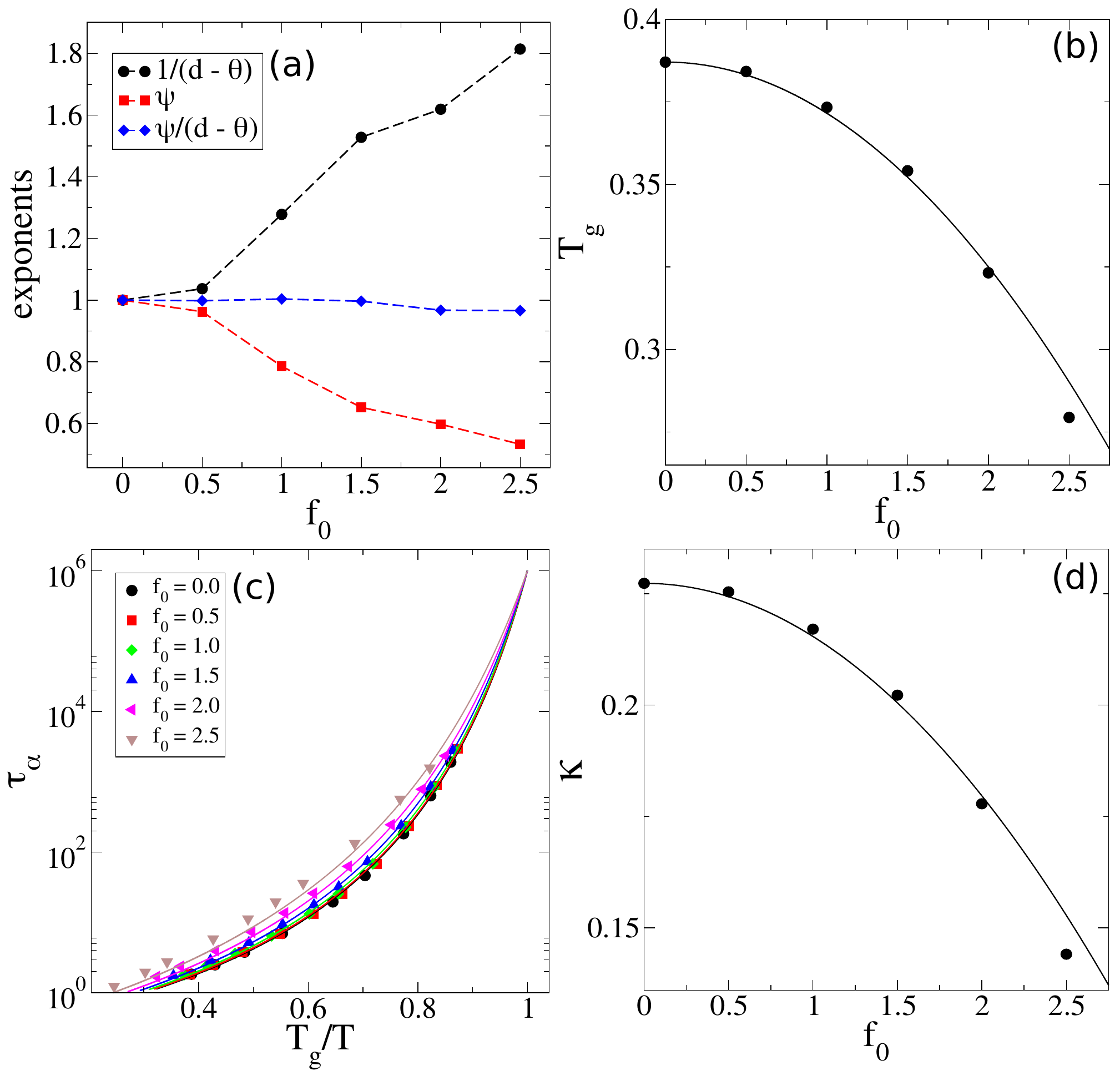}
	\caption{(a) For the relaxation dynamics, Eq. (\ref{taueq}), the combination $\psi/(d-\theta)$ is important. We find that although the individual exponents depend on activity, the combination remains almost independent of activity. (b) The experimental glass transition, $T_g$, as a function of $f_0$. (c) Angell plot for relaxation time, $\tau$, as a function of $T_g/T$ at different activity. (d) Kinetic fragility, $\kappa$, decreases as $f_0$ increases. Symbols represent simulation data, and the solid lines are theoretical predictions (not fits). }
	\label{angell_expo}
\end{figure}

The cornerstone of RFOT theory is the mosaic picture with the typical length scale $\xi_S$. The theory provides the expression of $\tau$ via the relaxation of a typical mosaic. Thus, the exponents $\theta$ and $\psi$ are pivotal within the theory. We have independently obtained these exponents, as shown in Figs. (\ref{figure1}d) and (\ref{psixitau}b), and find that activity affects both of them. Despite this nontrivial activity-dependence of the exponents, how the relaxation dynamics looks equilibrium-like, and how did the formulation of Ref. \cite{activerfot} provide good agreements with the simulation data? As evident from Eq. (\ref{taueq}), within RFOT theory, the combination, $\psi/(d-\theta)$, is the important parameter for the relaxation dynamics. Figure \ref{angell_expo}(a) shows this combination, together with the individual exponents. Despite the strong activity dependence of the individual parameters, where $\psi$ and $\theta$ changes almost linearly with $f_0$, the value of $\psi/(d-\theta)$ remains nearly constant and equal to one; the same value as Ref. \cite{activerfot}, which was motivated from the argument that the relaxation dynamics remains effectively equilibrium-like. This explains the effective equilibrium-like behavior of the relaxation dynamics.

Finally, we show some further comparisons of the theory with simulation data. Note that we have already obtained all the constants within the theory. Figure \ref{angell_expo}(b) shows the glass transition temperature, $T_g$, defined via $\tau(T_g)=10^6$, both in simulation and theory. In our simulations, we fit the data with the Vogel-Fulcher form $\tau=\tau_0\exp[E/(T-T_K)]$ and obtain $T_g$ using these parameters. The theory agrees quite well with the simulation data. Figure \ref{angell_expo}(c) shows the relaxation time data in the Angell plot representation. Figure \ref{angell_expo}(d) shows the kinetic fragility obtained via fitting the data of $\tau$ with the form $\log(\tau/\tau_0)=[\kappa(f_0)(T/T_K-1)]^{-1}$ and that from the theory. Note that {\em none} of these theoretical curves are fits; they are plots as obtained from Eq. (\ref{taueq}) with $\psi/(d-\theta)=1$ with the fitting parameters already fixed previously. These comparisons show that the theory agrees remarkably well with the simulation results.

The active RFOT theory presented here and in Ref. \cite{activerfot} is a phenomenological extension of the equilibrium theory for active systems. In principle, both the surface energy and configurational entropy terms should be affected by activity. However, comparison with simulation data shows the activity correction to the surface energy term is small, and we have ignored it for simplicity. The inclusion of this correction does not affect our qualitative results. We emphasize that the other nontrivial effect of activity on the surface energy, namely in the exponent $\theta$, is contained within the theory presented here and forms the basis of one of the main results of this work. The additional correction in the high-activity regime, proposed in Ref. \cite{rituparno2021}, does not seem to be important in the regime of our interest in this work.
We emphasize that activity leads to much higher values of $\xi_S$, as shown in Fig. \ref{figure1}, compared to equilibrium systems. The RFOT theory exponents, $\psi$ and $\theta$, depend on activity. The thermodynamic transition point, $T_K^{eff}$, gets shifted to lower $T$. In particular, one can go to a regime where $T_K^{eff}$ becomes negative and finds sub-Arrhenius behavior of $\tau$. However, this can happen at relatively higher densities, unlike equilibrium systems \cite{berthier2009,monojJPCB}. Sub-Arrhenius behavior is readily found in confluent systems \cite{souvik2021}. Whether the origins of such behavior in these distinctively different systems are similar or not remains unclear.

In conclusion, the results presented here show that although the relaxation dynamics in an active glass remains equilibrium-like, activity has nontrivial effects on the glassy properties. These nontrivial effects manifest via the activity dependence of the RFOT theory exponents, $\theta$ and $\psi$. But, their combination, $\psi/(d-\theta)$, becomes nearly independent of activity. This specific combination of the exponents controls the relaxation dynamics (Eq. \ref{taueq}). This explains the effectively equilibrium-like nature of the relaxation dynamics in active systems. In addition, we find the static length scale, $\xi_S$, in the presence of activity can grow much higher than in equilibrium. Given the frustratingly small values of $\xi_S$ in equilibrium systems, the nontrivial effects of activity along with larger $\xi_S$ should encourage devising more rigorous theories whose predictions can be easier to compare with simulations and experiments of active glasses.

\noindent{\bf Acknowledgments:} 
We acknowledge support of the Department of Atomic Energy, Government of India, under Project Identification 
No. RTI 4007. SK acknowledges support from Core Research Grant CRG/2019/005373 from Science and 
Engineering Research Board (SERB) as well as Swarna Jayanti Fellowship grants DST/SJF/PSA-01/2018-19 and 
SB/SFJ/2019-20/05. KP acknowledges financial support from SB/SFJ/2019-20/05.

\bibliography{activeRFOT_reference.bib}

\begin{thebibliography}{51}%
\makeatletter
\providecommand \@ifxundefined [1]{%
 \@ifx{#1\undefined}
}%
\providecommand \@ifnum [1]{%
 \ifnum #1\expandafter \@firstoftwo
 \else \expandafter \@secondoftwo
 \fi
}%
\providecommand \@ifx [1]{%
 \ifx #1\expandafter \@firstoftwo
 \else \expandafter \@secondoftwo
 \fi
}%
\providecommand \natexlab [1]{#1}%
\providecommand \enquote  [1]{``#1''}%
\providecommand \bibnamefont  [1]{#1}%
\providecommand \bibfnamefont [1]{#1}%
\providecommand \citenamefont [1]{#1}%
\providecommand \href@noop [0]{\@secondoftwo}%
\providecommand \href [0]{\begingroup \@sanitize@url \@href}%
\providecommand \@href[1]{\@@startlink{#1}\@@href}%
\providecommand \@@href[1]{\endgroup#1\@@endlink}%
\providecommand \@sanitize@url [0]{\catcode `\\12\catcode `\$12\catcode
  `\&12\catcode `\#12\catcode `\^12\catcode `\_12\catcode `\%12\relax}%
\providecommand \@@startlink[1]{}%
\providecommand \@@endlink[0]{}%
\providecommand \url  [0]{\begingroup\@sanitize@url \@url }%
\providecommand \@url [1]{\endgroup\@href {#1}{\urlprefix }}%
\providecommand \urlprefix  [0]{URL }%
\providecommand \Eprint [0]{\href }%
\providecommand \doibase [0]{http://dx.doi.org/}%
\providecommand \selectlanguage [0]{\@gobble}%
\providecommand \bibinfo  [0]{\@secondoftwo}%
\providecommand \bibfield  [0]{\@secondoftwo}%
\providecommand \translation [1]{[#1]}%
\providecommand \BibitemOpen [0]{}%
\providecommand \bibitemStop [0]{}%
\providecommand \bibitemNoStop [0]{.\EOS\space}%
\providecommand \EOS [0]{\spacefactor3000\relax}%
\providecommand \BibitemShut  [1]{\csname bibitem#1\endcsname}%
\let\auto@bib@innerbib\@empty
\bibitem [{\citenamefont {Berthier}\ and\ \citenamefont
  {Biroli}(2011)}]{giulioreview}%
  \BibitemOpen
  \bibfield  {author} {\bibinfo {author} {\bibfnamefont {L.}~\bibnamefont
  {Berthier}}\ and\ \bibinfo {author} {\bibfnamefont {G.}~\bibnamefont
  {Biroli}},\ }\href {\doibase 10.1103/RevModPhys.83.587} {\bibfield  {journal}
  {\bibinfo  {journal} {Rev. Mod. Phys.}\ }\textbf {\bibinfo {volume} {83}},\
  \bibinfo {pages} {587} (\bibinfo {year} {2011})}\BibitemShut {NoStop}%
\bibitem [{\citenamefont {Lubchenko}\ and\ \citenamefont
  {Wolynes}(2007)}]{lubchenko2007}%
  \BibitemOpen
  \bibfield  {author} {\bibinfo {author} {\bibfnamefont {V.}~\bibnamefont
  {Lubchenko}}\ and\ \bibinfo {author} {\bibfnamefont {P.~G.}\ \bibnamefont
  {Wolynes}},\ }\href {\doibase 10.1146/annurev.physchem.58.032806.104653}
  {\bibfield  {journal} {\bibinfo  {journal} {Annu. Rev. Phys. Chem.}\ }\textbf
  {\bibinfo {volume} {58}},\ \bibinfo {pages} {235} (\bibinfo {year}
  {2007})}\BibitemShut {NoStop}%
\bibitem [{\citenamefont {Das}(2004)}]{das2004}%
  \BibitemOpen
  \bibfield  {author} {\bibinfo {author} {\bibfnamefont {S.~P.}\ \bibnamefont
  {Das}},\ }\href {\doibase 10.1103/RevModPhys.76.785} {\bibfield  {journal}
  {\bibinfo  {journal} {Rev. Mod. Phys.}\ }\textbf {\bibinfo {volume} {76}},\
  \bibinfo {pages} {785} (\bibinfo {year} {2004})}\BibitemShut {NoStop}%
\bibitem [{\citenamefont {Dyre}\ \emph {et~al.}(2009)\citenamefont {Dyre},
  \citenamefont {Hechsher},\ and\ \citenamefont {Niss}}]{dyre2009}%
  \BibitemOpen
  \bibfield  {author} {\bibinfo {author} {\bibfnamefont {J.~C.}\ \bibnamefont
  {Dyre}}, \bibinfo {author} {\bibfnamefont {T.}~\bibnamefont {Hechsher}}, \
  and\ \bibinfo {author} {\bibfnamefont {K.}~\bibnamefont {Niss}},\ }\href
  {\doibase 10.1016/j.jnoncrysol.2009.01.039} {\bibfield  {journal} {\bibinfo
  {journal} {J. Non-Cryst. Solids}\ }\textbf {\bibinfo {volume} {355}},\
  \bibinfo {pages} {624} (\bibinfo {year} {2009})}\BibitemShut {NoStop}%
\bibitem [{\citenamefont {Biroli}\ and\ \citenamefont
  {Bouchaud}(2012)}]{giuliobook}%
  \BibitemOpen
  \bibfield  {author} {\bibinfo {author} {\bibfnamefont {G.}~\bibnamefont
  {Biroli}}\ and\ \bibinfo {author} {\bibfnamefont {J.~P.}\ \bibnamefont
  {Bouchaud}},\ }in\ \href@noop {} {\emph {\bibinfo {booktitle} {Structural
  Glasses and Supercooled Liquids: Theory, Experiment, and Applications}}},\
  \bibinfo {editor} {edited by\ \bibinfo {editor} {\bibfnamefont {P.~G.}\
  \bibnamefont {Wolynes}}\ and\ \bibinfo {editor} {\bibfnamefont
  {V.}~\bibnamefont {Lubchenko}}}\ (\bibinfo  {publisher} {John Wiley and Sons,
  Inc., Hoboken, NJ, USA},\ \bibinfo {year} {2012})\BibitemShut {NoStop}%
\bibitem [{\citenamefont {Tarjus}\ \emph {et~al.}(2005)\citenamefont {Tarjus},
  \citenamefont {Kivelson}, \citenamefont {Nussinov},\ and\ \citenamefont
  {Viot}}]{tarjus2005}%
  \BibitemOpen
  \bibfield  {author} {\bibinfo {author} {\bibfnamefont {G.}~\bibnamefont
  {Tarjus}}, \bibinfo {author} {\bibfnamefont {S.~A.}\ \bibnamefont
  {Kivelson}}, \bibinfo {author} {\bibfnamefont {Z.}~\bibnamefont {Nussinov}},
  \ and\ \bibinfo {author} {\bibfnamefont {P.}~\bibnamefont {Viot}},\ }\href
  {\doibase 10.1088/0953-8984/17/50/R01} {\bibfield  {journal} {\bibinfo
  {journal} {J. Phys.: Condens. Matter}\ }\textbf {\bibinfo {volume} {17}},\
  \bibinfo {pages} {R1143} (\bibinfo {year} {2005})}\BibitemShut {NoStop}%
\bibitem [{\citenamefont {Karmakar}\ \emph {et~al.}(2014)\citenamefont
  {Karmakar}, \citenamefont {Dasgupta},\ and\ \citenamefont
  {Sastry}}]{smarajitAnnual}%
  \BibitemOpen
  \bibfield  {author} {\bibinfo {author} {\bibfnamefont {S.}~\bibnamefont
  {Karmakar}}, \bibinfo {author} {\bibfnamefont {C.}~\bibnamefont {Dasgupta}},
  \ and\ \bibinfo {author} {\bibfnamefont {S.}~\bibnamefont {Sastry}},\ }\href
  {\doibase 10.1146/annurev-conmatphys-031113-133848} {\bibfield  {journal}
  {\bibinfo  {journal} {Annual Review of Condensed Matter Physics}\ }\textbf
  {\bibinfo {volume} {5}},\ \bibinfo {pages} {255} (\bibinfo {year}
  {2014})}\BibitemShut {NoStop}%
\bibitem [{\citenamefont {Ramaswamy}(2010)}]{sriramreview}%
  \BibitemOpen
  \bibfield  {author} {\bibinfo {author} {\bibfnamefont {S.}~\bibnamefont
  {Ramaswamy}},\ }\href {\doibase 10.1146/annurev-conmatphys-070909-104101}
  {\bibfield  {journal} {\bibinfo  {journal} {Annu. Rev. Condens. Matter
  Phys.}\ }\textbf {\bibinfo {volume} {1}},\ \bibinfo {pages} {323} (\bibinfo
  {year} {2010})}\BibitemShut {NoStop}%
\bibitem [{\citenamefont {Marchetti}\ \emph {et~al.}(2013)\citenamefont
  {Marchetti}, \citenamefont {Joanny}, \citenamefont {Ramaswamy}, \citenamefont
  {Liverpool}, \citenamefont {Prost}, \citenamefont {Rao},\ and\ \citenamefont
  {Simha}}]{sriramrmp}%
  \BibitemOpen
  \bibfield  {author} {\bibinfo {author} {\bibfnamefont {M.~C.}\ \bibnamefont
  {Marchetti}}, \bibinfo {author} {\bibfnamefont {J.~F.}\ \bibnamefont
  {Joanny}}, \bibinfo {author} {\bibfnamefont {S.}~\bibnamefont {Ramaswamy}},
  \bibinfo {author} {\bibfnamefont {T.~B.}\ \bibnamefont {Liverpool}}, \bibinfo
  {author} {\bibfnamefont {J.}~\bibnamefont {Prost}}, \bibinfo {author}
  {\bibfnamefont {M.}~\bibnamefont {Rao}}, \ and\ \bibinfo {author}
  {\bibfnamefont {R.~A.}\ \bibnamefont {Simha}},\ }\href {\doibase
  10.1103/RevModPhys.85.1143} {\bibfield  {journal} {\bibinfo  {journal} {Rev.
  Mod. Phys.}\ }\textbf {\bibinfo {volume} {85}},\ \bibinfo {pages} {1143}
  (\bibinfo {year} {2013})}\BibitemShut {NoStop}%
\bibitem [{\citenamefont {Poujade}\ \emph {et~al.}(2007)\citenamefont
  {Poujade}, \citenamefont {Grasland-Mongrain}, \citenamefont {Hertzog},
  \citenamefont {Jouanneau}, \citenamefont {Chavrier}, \citenamefont {Ladoux},
  \citenamefont {Buguin},\ and\ \citenamefont {Silberzan}}]{poujade2007}%
  \BibitemOpen
  \bibfield  {author} {\bibinfo {author} {\bibfnamefont {M.}~\bibnamefont
  {Poujade}}, \bibinfo {author} {\bibfnamefont {E.}~\bibnamefont
  {Grasland-Mongrain}}, \bibinfo {author} {\bibfnamefont {A.}~\bibnamefont
  {Hertzog}}, \bibinfo {author} {\bibfnamefont {J.}~\bibnamefont {Jouanneau}},
  \bibinfo {author} {\bibfnamefont {P.}~\bibnamefont {Chavrier}}, \bibinfo
  {author} {\bibfnamefont {B.}~\bibnamefont {Ladoux}}, \bibinfo {author}
  {\bibfnamefont {A.}~\bibnamefont {Buguin}}, \ and\ \bibinfo {author}
  {\bibfnamefont {P.}~\bibnamefont {Silberzan}},\ }\href {\doibase
  10.1073/pnas.0705062104} {\bibfield  {journal} {\bibinfo  {journal} {Proc.
  Natl. Acad. Sci. (USA)}\ }\textbf {\bibinfo {volume} {104}},\ \bibinfo
  {pages} {15988} (\bibinfo {year} {2007})}\BibitemShut {NoStop}%
\bibitem [{\citenamefont {Cochet-Escartin}\ \emph {et~al.}(2014)\citenamefont
  {Cochet-Escartin}, \citenamefont {Ranft}, \citenamefont {Silberzan},\ and\
  \citenamefont {Marcq}}]{olivier2014}%
  \BibitemOpen
  \bibfield  {author} {\bibinfo {author} {\bibfnamefont {O.}~\bibnamefont
  {Cochet-Escartin}}, \bibinfo {author} {\bibfnamefont {J.}~\bibnamefont
  {Ranft}}, \bibinfo {author} {\bibfnamefont {P.}~\bibnamefont {Silberzan}}, \
  and\ \bibinfo {author} {\bibfnamefont {P.}~\bibnamefont {Marcq}},\ }\href
  {\doibase 10.1016/j.bpj.2013.11.015} {\bibfield  {journal} {\bibinfo
  {journal} {Biophys. J.}\ }\textbf {\bibinfo {volume} {106}},\ \bibinfo
  {pages} {65} (\bibinfo {year} {2014})}\BibitemShut {NoStop}%
\bibitem [{\citenamefont {Angelini}\ \emph {et~al.}(2011)\citenamefont
  {Angelini}, \citenamefont {Hannezo}, \citenamefont {Trepat}, \citenamefont
  {Marquez}, \citenamefont {Fredberg},\ and\ \citenamefont
  {Weitz}}]{angelini2011}%
  \BibitemOpen
  \bibfield  {author} {\bibinfo {author} {\bibfnamefont {T.~E.}\ \bibnamefont
  {Angelini}}, \bibinfo {author} {\bibfnamefont {E.}~\bibnamefont {Hannezo}},
  \bibinfo {author} {\bibfnamefont {X.}~\bibnamefont {Trepat}}, \bibinfo
  {author} {\bibfnamefont {M.}~\bibnamefont {Marquez}}, \bibinfo {author}
  {\bibfnamefont {J.~J.}\ \bibnamefont {Fredberg}}, \ and\ \bibinfo {author}
  {\bibfnamefont {D.~A.}\ \bibnamefont {Weitz}},\ }\href {\doibase
  10.1073/pnas.1010059108} {\bibfield  {journal} {\bibinfo  {journal} {Proc.
  Natl. Acad. Sci. (USA)}\ }\textbf {\bibinfo {volume} {108}},\ \bibinfo
  {pages} {4717} (\bibinfo {year} {2011})}\BibitemShut {NoStop}%
\bibitem [{\citenamefont {Park}\ \emph {et~al.}(2015)\citenamefont {Park},
  \citenamefont {Kim}, \citenamefont {Bi}, \citenamefont {Mitchel},
  \citenamefont {Qazvini}, \citenamefont {Tantisira}, \citenamefont {Park},
  \citenamefont {McGill}, \citenamefont {Kim}, \citenamefont {Gweon},
  \citenamefont {Notbohm}, \citenamefont {Jr}, \citenamefont {Burger},
  \citenamefont {Randell}, \citenamefont {Kho}, \citenamefont {Tambe},
  \citenamefont {Hardin}, \citenamefont {Shore}, \citenamefont {Israel},
  \citenamefont {Weitz}, \citenamefont {Tschumperlin}, \citenamefont {Henske},
  \citenamefont {Weiss}, \citenamefont {Manning}, \citenamefont {Butler},
  \citenamefont {Drazen},\ and\ \citenamefont {Fredberg}}]{park2015}%
  \BibitemOpen
  \bibfield  {author} {\bibinfo {author} {\bibfnamefont {J.-A.}\ \bibnamefont
  {Park}}, \bibinfo {author} {\bibfnamefont {J.~H.}\ \bibnamefont {Kim}},
  \bibinfo {author} {\bibfnamefont {D.}~\bibnamefont {Bi}}, \bibinfo {author}
  {\bibfnamefont {J.~A.}\ \bibnamefont {Mitchel}}, \bibinfo {author}
  {\bibfnamefont {N.~T.}\ \bibnamefont {Qazvini}}, \bibinfo {author}
  {\bibfnamefont {K.}~\bibnamefont {Tantisira}}, \bibinfo {author}
  {\bibfnamefont {C.~Y.}\ \bibnamefont {Park}}, \bibinfo {author}
  {\bibfnamefont {M.}~\bibnamefont {McGill}}, \bibinfo {author} {\bibfnamefont
  {S.-H.}\ \bibnamefont {Kim}}, \bibinfo {author} {\bibfnamefont
  {B.}~\bibnamefont {Gweon}}, \bibinfo {author} {\bibfnamefont
  {J.}~\bibnamefont {Notbohm}}, \bibinfo {author} {\bibfnamefont {R.~S.}\
  \bibnamefont {Jr}}, \bibinfo {author} {\bibfnamefont {S.}~\bibnamefont
  {Burger}}, \bibinfo {author} {\bibfnamefont {S.~H.}\ \bibnamefont {Randell}},
  \bibinfo {author} {\bibfnamefont {A.~T.}\ \bibnamefont {Kho}}, \bibinfo
  {author} {\bibfnamefont {D.~T.}\ \bibnamefont {Tambe}}, \bibinfo {author}
  {\bibfnamefont {C.}~\bibnamefont {Hardin}}, \bibinfo {author} {\bibfnamefont
  {S.~A.}\ \bibnamefont {Shore}}, \bibinfo {author} {\bibfnamefont
  {E.}~\bibnamefont {Israel}}, \bibinfo {author} {\bibfnamefont {D.~A.}\
  \bibnamefont {Weitz}}, \bibinfo {author} {\bibfnamefont {D.~J.}\ \bibnamefont
  {Tschumperlin}}, \bibinfo {author} {\bibfnamefont {E.~P.}\ \bibnamefont
  {Henske}}, \bibinfo {author} {\bibfnamefont {S.~T.}\ \bibnamefont {Weiss}},
  \bibinfo {author} {\bibfnamefont {M.~L.}\ \bibnamefont {Manning}}, \bibinfo
  {author} {\bibfnamefont {J.~P.}\ \bibnamefont {Butler}}, \bibinfo {author}
  {\bibfnamefont {J.~M.}\ \bibnamefont {Drazen}}, \ and\ \bibinfo {author}
  {\bibfnamefont {J.~J.}\ \bibnamefont {Fredberg}},\ }\href {\doibase
  10.1038/NMAT4357} {\bibfield  {journal} {\bibinfo  {journal} {Nat. Mat.}\
  }\textbf {\bibinfo {volume} {14}},\ \bibinfo {pages} {1040} (\bibinfo {year}
  {2015})}\BibitemShut {NoStop}%
\bibitem [{\citenamefont {Malinverno}\ \emph {et~al.}(2017)\citenamefont
  {Malinverno}, \citenamefont {Corallino}, \citenamefont {Giavazzi},
  \citenamefont {Bergert}, \citenamefont {Li}, \citenamefont {Leoni},
  \citenamefont {Disanza}, \citenamefont {Frittoli}, \citenamefont {Oldani},
  \citenamefont {Martini}, \citenamefont {Lendenmann}, \citenamefont
  {Deflorian}, \citenamefont {Beznoussenko}, \citenamefont {Poulikakos},
  \citenamefont {Ong}, \citenamefont {Uroz}, \citenamefont {Trepat},
  \citenamefont {Parazzoli}, \citenamefont {Maiuri}, \citenamefont {Yu},
  \citenamefont {Ferrari}, \citenamefont {Cerbino},\ and\ \citenamefont
  {Scita}}]{malinverno2017}%
  \BibitemOpen
  \bibfield  {author} {\bibinfo {author} {\bibfnamefont {C.}~\bibnamefont
  {Malinverno}}, \bibinfo {author} {\bibfnamefont {S.}~\bibnamefont
  {Corallino}}, \bibinfo {author} {\bibfnamefont {F.}~\bibnamefont {Giavazzi}},
  \bibinfo {author} {\bibfnamefont {M.}~\bibnamefont {Bergert}}, \bibinfo
  {author} {\bibfnamefont {Q.}~\bibnamefont {Li}}, \bibinfo {author}
  {\bibfnamefont {M.}~\bibnamefont {Leoni}}, \bibinfo {author} {\bibfnamefont
  {A.}~\bibnamefont {Disanza}}, \bibinfo {author} {\bibfnamefont
  {E.}~\bibnamefont {Frittoli}}, \bibinfo {author} {\bibfnamefont
  {A.}~\bibnamefont {Oldani}}, \bibinfo {author} {\bibfnamefont
  {E.}~\bibnamefont {Martini}}, \bibinfo {author} {\bibfnamefont
  {T.}~\bibnamefont {Lendenmann}}, \bibinfo {author} {\bibfnamefont
  {G.}~\bibnamefont {Deflorian}}, \bibinfo {author} {\bibfnamefont {G.~V.}\
  \bibnamefont {Beznoussenko}}, \bibinfo {author} {\bibfnamefont
  {D.}~\bibnamefont {Poulikakos}}, \bibinfo {author} {\bibfnamefont {K.~H.}\
  \bibnamefont {Ong}}, \bibinfo {author} {\bibfnamefont {M.}~\bibnamefont
  {Uroz}}, \bibinfo {author} {\bibfnamefont {X.}~\bibnamefont {Trepat}},
  \bibinfo {author} {\bibfnamefont {D.}~\bibnamefont {Parazzoli}}, \bibinfo
  {author} {\bibfnamefont {P.}~\bibnamefont {Maiuri}}, \bibinfo {author}
  {\bibfnamefont {W.}~\bibnamefont {Yu}}, \bibinfo {author} {\bibfnamefont
  {A.}~\bibnamefont {Ferrari}}, \bibinfo {author} {\bibfnamefont
  {R.}~\bibnamefont {Cerbino}}, \ and\ \bibinfo {author} {\bibfnamefont
  {G.}~\bibnamefont {Scita}},\ }\href {\doibase 10.1038/NMAT4848} {\bibfield
  {journal} {\bibinfo  {journal} {Nat. Mat.}\ }\textbf {\bibinfo {volume}
  {16}},\ \bibinfo {pages} {587} (\bibinfo {year} {2017})}\BibitemShut
  {NoStop}%
\bibitem [{\citenamefont {Garcia}\ \emph {et~al.}(2015)\citenamefont {Garcia},
  \citenamefont {Hannezo}, \citenamefont {Elgeti}, \citenamefont {Joanny},
  \citenamefont {Silberzan},\ and\ \citenamefont {Gov}}]{garcia2015}%
  \BibitemOpen
  \bibfield  {author} {\bibinfo {author} {\bibfnamefont {S.}~\bibnamefont
  {Garcia}}, \bibinfo {author} {\bibfnamefont {E.}~\bibnamefont {Hannezo}},
  \bibinfo {author} {\bibfnamefont {J.}~\bibnamefont {Elgeti}}, \bibinfo
  {author} {\bibfnamefont {J.~F.}\ \bibnamefont {Joanny}}, \bibinfo {author}
  {\bibfnamefont {P.}~\bibnamefont {Silberzan}}, \ and\ \bibinfo {author}
  {\bibfnamefont {N.~S.}\ \bibnamefont {Gov}},\ }\href {\doibase
  10.1073/pnas.1510973112} {\bibfield  {journal} {\bibinfo  {journal} {Proc.
  Natl. Acad. Sci. (USA)}\ }\textbf {\bibinfo {volume} {112}},\ \bibinfo
  {pages} {15314} (\bibinfo {year} {2015})}\BibitemShut {NoStop}%
\bibitem [{\citenamefont {Parry}\ \emph {et~al.}(2014)\citenamefont {Parry},
  \citenamefont {Surovtsev}, \citenamefont {Cabenn}, \citenamefont {O'Hern},
  \citenamefont {Dufresne},\ and\ \citenamefont {Jacobs-Wagner}}]{parry2014}%
  \BibitemOpen
  \bibfield  {author} {\bibinfo {author} {\bibfnamefont {B.~R.}\ \bibnamefont
  {Parry}}, \bibinfo {author} {\bibfnamefont {I.~V.}\ \bibnamefont
  {Surovtsev}}, \bibinfo {author} {\bibfnamefont {M.~T.}\ \bibnamefont
  {Cabenn}}, \bibinfo {author} {\bibfnamefont {C.~S.}\ \bibnamefont {O'Hern}},
  \bibinfo {author} {\bibfnamefont {E.~R.}\ \bibnamefont {Dufresne}}, \ and\
  \bibinfo {author} {\bibfnamefont {C.}~\bibnamefont {Jacobs-Wagner}},\ }\href
  {\doibase 10.1016/j.cell.2013.11.028} {\bibfield  {journal} {\bibinfo
  {journal} {Cell}\ }\textbf {\bibinfo {volume} {156}},\ \bibinfo {pages} {183}
  (\bibinfo {year} {2014})}\BibitemShut {NoStop}%
\bibitem [{\citenamefont {Nishizawa}\ \emph {et~al.}(2017)\citenamefont
  {Nishizawa}, \citenamefont {Fujiwara}, \citenamefont {Ikenaga}, \citenamefont
  {Nakajo}, \citenamefont {Yanagisawa},\ and\ \citenamefont
  {Mizuno}}]{nishizawa2017}%
  \BibitemOpen
  \bibfield  {author} {\bibinfo {author} {\bibfnamefont {K.}~\bibnamefont
  {Nishizawa}}, \bibinfo {author} {\bibfnamefont {K.}~\bibnamefont {Fujiwara}},
  \bibinfo {author} {\bibfnamefont {M.}~\bibnamefont {Ikenaga}}, \bibinfo
  {author} {\bibfnamefont {N.}~\bibnamefont {Nakajo}}, \bibinfo {author}
  {\bibfnamefont {M.}~\bibnamefont {Yanagisawa}}, \ and\ \bibinfo {author}
  {\bibfnamefont {D.}~\bibnamefont {Mizuno}},\ }\href {\doibase
  10.1038/s41598-017-14883-y} {\bibfield  {journal} {\bibinfo  {journal} {Sci.
  Rep.}\ }\textbf {\bibinfo {volume} {7}},\ \bibinfo {pages} {15143} (\bibinfo
  {year} {2017})}\BibitemShut {NoStop}%
\bibitem [{\citenamefont {Zhou}\ \emph {et~al.}(2009)\citenamefont {Zhou},
  \citenamefont {Trepat}, \citenamefont {Park}, \citenamefont {Lenormand},
  \citenamefont {Oliver}, \citenamefont {Mijailovich}, \citenamefont {Hardin},
  \citenamefont {Weitz}, \citenamefont {Butler},\ and\ \citenamefont
  {Fredberg}}]{zhou2009}%
  \BibitemOpen
  \bibfield  {author} {\bibinfo {author} {\bibfnamefont {E.~H.}\ \bibnamefont
  {Zhou}}, \bibinfo {author} {\bibfnamefont {X.}~\bibnamefont {Trepat}},
  \bibinfo {author} {\bibfnamefont {C.~Y.}\ \bibnamefont {Park}}, \bibinfo
  {author} {\bibfnamefont {G.}~\bibnamefont {Lenormand}}, \bibinfo {author}
  {\bibfnamefont {M.~N.}\ \bibnamefont {Oliver}}, \bibinfo {author}
  {\bibfnamefont {S.~M.}\ \bibnamefont {Mijailovich}}, \bibinfo {author}
  {\bibfnamefont {C.}~\bibnamefont {Hardin}}, \bibinfo {author} {\bibfnamefont
  {D.~A.}\ \bibnamefont {Weitz}}, \bibinfo {author} {\bibfnamefont {J.~P.}\
  \bibnamefont {Butler}}, \ and\ \bibinfo {author} {\bibfnamefont {J.~J.}\
  \bibnamefont {Fredberg}},\ }\href {\doibase 10.1073pnas.0901462106}
  {\bibfield  {journal} {\bibinfo  {journal} {Proc. Natl. Acad. Sci. (USA)}\
  }\textbf {\bibinfo {volume} {106}},\ \bibinfo {pages} {10632} (\bibinfo
  {year} {2009})}\BibitemShut {NoStop}%
\bibitem [{\citenamefont {Malmi-Kakkada}\ \emph {et~al.}(2018)\citenamefont
  {Malmi-Kakkada}, \citenamefont {Li}, \citenamefont {Samanta}, \citenamefont
  {Sinha},\ and\ \citenamefont {Thirumalai}}]{kakkada2018}%
  \BibitemOpen
  \bibfield  {author} {\bibinfo {author} {\bibfnamefont {A.~N.}\ \bibnamefont
  {Malmi-Kakkada}}, \bibinfo {author} {\bibfnamefont {X.}~\bibnamefont {Li}},
  \bibinfo {author} {\bibfnamefont {H.~S.}\ \bibnamefont {Samanta}}, \bibinfo
  {author} {\bibfnamefont {S.}~\bibnamefont {Sinha}}, \ and\ \bibinfo {author}
  {\bibfnamefont {D.}~\bibnamefont {Thirumalai}},\ }\href {\doibase
  10.1103/PhysRevX.8.021025} {\bibfield  {journal} {\bibinfo  {journal} {Phys.
  Rev. X}\ }\textbf {\bibinfo {volume} {8}},\ \bibinfo {pages} {021025}
  (\bibinfo {year} {2018})}\BibitemShut {NoStop}%
\bibitem [{\citenamefont {Streitberger}\ \emph {et~al.}(2020)\citenamefont
  {Streitberger}, \citenamefont {Lilaj}, \citenamefont {Schrank}, \citenamefont
  {J{\"{u}}rgen~Braun}, \citenamefont {Reiss-Zimmermann}, \citenamefont
  {K{\"{a}}s},\ and\ \citenamefont {Sack}}]{streitberger2020}%
  \BibitemOpen
  \bibfield  {author} {\bibinfo {author} {\bibfnamefont {K.-J.}\ \bibnamefont
  {Streitberger}}, \bibinfo {author} {\bibfnamefont {L.}~\bibnamefont {Lilaj}},
  \bibinfo {author} {\bibfnamefont {F.}~\bibnamefont {Schrank}}, \bibinfo
  {author} {\bibfnamefont {a.~K.-T.~H.}\ \bibnamefont {J{\"{u}}rgen~Braun}},
  \bibinfo {author} {\bibfnamefont {M.}~\bibnamefont {Reiss-Zimmermann}},
  \bibinfo {author} {\bibfnamefont {J.~A.}\ \bibnamefont {K{\"{a}}s}}, \ and\
  \bibinfo {author} {\bibfnamefont {I.}~\bibnamefont {Sack}},\ }\href {\doibase
  10.1073/pnas.1913511116} {\bibfield  {journal} {\bibinfo  {journal} {Proc.
  Natl. Acad. Sci. (USA)}\ }\textbf {\bibinfo {volume} {117}},\ \bibinfo
  {pages} {128} (\bibinfo {year} {2020})}\BibitemShut {NoStop}%
\bibitem [{\citenamefont {Berthier}(2014)}]{berthier2014}%
  \BibitemOpen
  \bibfield  {author} {\bibinfo {author} {\bibfnamefont {L.}~\bibnamefont
  {Berthier}},\ }\href {\doibase 10.1103/PhysRevLett.112.220602} {\bibfield
  {journal} {\bibinfo  {journal} {Phys. Rev. Lett.}\ }\textbf {\bibinfo
  {volume} {112}},\ \bibinfo {pages} {220602} (\bibinfo {year}
  {2014})}\BibitemShut {NoStop}%
\bibitem [{\citenamefont {Paul}\ \emph {et~al.}(2021)\citenamefont {Paul},
  \citenamefont {Nandi},\ and\ \citenamefont {Karmakar}}]{kallol2021}%
  \BibitemOpen
  \bibfield  {author} {\bibinfo {author} {\bibfnamefont {K.}~\bibnamefont
  {Paul}}, \bibinfo {author} {\bibfnamefont {S.~K.}\ \bibnamefont {Nandi}}, \
  and\ \bibinfo {author} {\bibfnamefont {S.}~\bibnamefont {Karmakar}},\ }\href
  {https://arxiv.org/abs/2105.12702} {\bibfield  {journal} {\bibinfo  {journal}
  {arXiv: 2105.12702}\ } (\bibinfo {year} {2021})}\BibitemShut {NoStop}%
\bibitem [{\citenamefont {Mandal}\ \emph {et~al.}(2016)\citenamefont {Mandal},
  \citenamefont {Bhuyan}, \citenamefont {Rao},\ and\ \citenamefont
  {Dasgupta}}]{mandal2016}%
  \BibitemOpen
  \bibfield  {author} {\bibinfo {author} {\bibfnamefont {R.}~\bibnamefont
  {Mandal}}, \bibinfo {author} {\bibfnamefont {P.~J.}\ \bibnamefont {Bhuyan}},
  \bibinfo {author} {\bibfnamefont {M.}~\bibnamefont {Rao}}, \ and\ \bibinfo
  {author} {\bibfnamefont {C.}~\bibnamefont {Dasgupta}},\ }\href@noop {}
  {\bibfield  {journal} {\bibinfo  {journal} {Soft Matter}\ }\textbf {\bibinfo
  {volume} {12}},\ \bibinfo {pages} {6268} (\bibinfo {year}
  {2016})}\BibitemShut {NoStop}%
\bibitem [{\citenamefont {Flenner}\ \emph {et~al.}(2016)\citenamefont
  {Flenner}, \citenamefont {Szamel},\ and\ \citenamefont
  {Berthier}}]{flenner2016}%
  \BibitemOpen
  \bibfield  {author} {\bibinfo {author} {\bibfnamefont {E.}~\bibnamefont
  {Flenner}}, \bibinfo {author} {\bibfnamefont {G.}~\bibnamefont {Szamel}}, \
  and\ \bibinfo {author} {\bibfnamefont {L.}~\bibnamefont {Berthier}},\ }\href
  {\doibase 10.1039/c6sm01322h} {\bibfield  {journal} {\bibinfo  {journal}
  {Soft Matter}\ }\textbf {\bibinfo {volume} {12}},\ \bibinfo {pages} {7136}
  (\bibinfo {year} {2016})}\BibitemShut {NoStop}%
\bibitem [{\citenamefont {Nandi}\ and\ \citenamefont {Gov}(2017)}]{activemct}%
  \BibitemOpen
  \bibfield  {author} {\bibinfo {author} {\bibfnamefont {S.~K.}\ \bibnamefont
  {Nandi}}\ and\ \bibinfo {author} {\bibfnamefont {N.~S.}\ \bibnamefont
  {Gov}},\ }\href {\doibase 10.1039/C7SM01648D} {\bibfield  {journal} {\bibinfo
   {journal} {Soft Matter}\ }\textbf {\bibinfo {volume} {13}},\ \bibinfo
  {pages} {7609} (\bibinfo {year} {2017})}\BibitemShut {NoStop}%
\bibitem [{\citenamefont {Berthier}\ and\ \citenamefont
  {Kurchan}(2013)}]{berthier2013}%
  \BibitemOpen
  \bibfield  {author} {\bibinfo {author} {\bibfnamefont {L.}~\bibnamefont
  {Berthier}}\ and\ \bibinfo {author} {\bibfnamefont {J.}~\bibnamefont
  {Kurchan}},\ }\href {\doibase 10.1038/nphys2592} {\bibfield  {journal}
  {\bibinfo  {journal} {Nat. Phys.}\ }\textbf {\bibinfo {volume} {9}},\
  \bibinfo {pages} {310} (\bibinfo {year} {2013})}\BibitemShut {NoStop}%
\bibitem [{\citenamefont {Feng}\ and\ \citenamefont {Hou}(2017)}]{feng2017}%
  \BibitemOpen
  \bibfield  {author} {\bibinfo {author} {\bibfnamefont {M.}~\bibnamefont
  {Feng}}\ and\ \bibinfo {author} {\bibfnamefont {Z.}~\bibnamefont {Hou}},\
  }\href {\doibase 10.1039/C7SM00852J} {\bibfield  {journal} {\bibinfo
  {journal} {Soft Matter}\ }\textbf {\bibinfo {volume} {13}},\ \bibinfo {pages}
  {4464} (\bibinfo {year} {2017})}\BibitemShut {NoStop}%
\bibitem [{\citenamefont {Szamel}(2016)}]{szamel2016}%
  \BibitemOpen
  \bibfield  {author} {\bibinfo {author} {\bibfnamefont {G.}~\bibnamefont
  {Szamel}},\ }\href {\doibase 10.1103/PhysRevE.93.012603} {\bibfield
  {journal} {\bibinfo  {journal} {Phys. Rev. E}\ }\textbf {\bibinfo {volume}
  {93}},\ \bibinfo {pages} {012603} (\bibinfo {year} {2016})}\BibitemShut
  {NoStop}%
\bibitem [{\citenamefont {Nandi}\ \emph {et~al.}(2018)\citenamefont {Nandi},
  \citenamefont {Mandal}, \citenamefont {Bhuyan}, \citenamefont {Dasgupta},
  \citenamefont {Rao},\ and\ \citenamefont {Gov}}]{activerfot}%
  \BibitemOpen
  \bibfield  {author} {\bibinfo {author} {\bibfnamefont {S.~K.}\ \bibnamefont
  {Nandi}}, \bibinfo {author} {\bibfnamefont {R.}~\bibnamefont {Mandal}},
  \bibinfo {author} {\bibfnamefont {P.~J.}\ \bibnamefont {Bhuyan}}, \bibinfo
  {author} {\bibfnamefont {C.}~\bibnamefont {Dasgupta}}, \bibinfo {author}
  {\bibfnamefont {M.}~\bibnamefont {Rao}}, \ and\ \bibinfo {author}
  {\bibfnamefont {N.~S.}\ \bibnamefont {Gov}},\ }\href {\doibase
  10.1073/pnas.1721324115} {\bibfield  {journal} {\bibinfo  {journal} {Proc.
  Natl. Acad. Sci. (USA)}\ }\textbf {\bibinfo {volume} {115}},\ \bibinfo
  {pages} {7688} (\bibinfo {year} {2018})}\BibitemShut {NoStop}%
\bibitem [{\citenamefont {Ni}\ \emph {et~al.}(2013)\citenamefont {Ni},
  \citenamefont {Stuart},\ and\ \citenamefont {Dijkstra}}]{ni2013}%
  \BibitemOpen
  \bibfield  {author} {\bibinfo {author} {\bibfnamefont {R.}~\bibnamefont
  {Ni}}, \bibinfo {author} {\bibfnamefont {M.~A.~C.}\ \bibnamefont {Stuart}}, \
  and\ \bibinfo {author} {\bibfnamefont {M.}~\bibnamefont {Dijkstra}},\ }\href
  {\doibase 10.1038/ncomms3704} {\bibfield  {journal} {\bibinfo  {journal}
  {Nat. Commun}\ }\textbf {\bibinfo {volume} {4}},\ \bibinfo {pages} {2704}
  (\bibinfo {year} {2013})}\BibitemShut {NoStop}%
\bibitem [{\citenamefont {Vicsek}\ \emph {et~al.}(1995)\citenamefont {Vicsek},
  \citenamefont {Czir{\'{o}}k}, \citenamefont {Ben-Jacob}, \citenamefont
  {Cohen},\ and\ \citenamefont {Shochet}}]{vicsek1995}%
  \BibitemOpen
  \bibfield  {author} {\bibinfo {author} {\bibfnamefont {T.}~\bibnamefont
  {Vicsek}}, \bibinfo {author} {\bibfnamefont {A.}~\bibnamefont
  {Czir{\'{o}}k}}, \bibinfo {author} {\bibfnamefont {E.}~\bibnamefont
  {Ben-Jacob}}, \bibinfo {author} {\bibfnamefont {I.}~\bibnamefont {Cohen}}, \
  and\ \bibinfo {author} {\bibfnamefont {O.}~\bibnamefont {Shochet}},\ }\href
  {\doibase 10.1103/PhysRevLett.75.1226} {\bibfield  {journal} {\bibinfo
  {journal} {Phys. Rev. Lett.}\ }\textbf {\bibinfo {volume} {75}},\ \bibinfo
  {pages} {1226} (\bibinfo {year} {1995})}\BibitemShut {NoStop}%
\bibitem [{\citenamefont {Narayan}\ \emph {et~al.}(2007)\citenamefont
  {Narayan}, \citenamefont {Ramaswamy},\ and\ \citenamefont
  {Menon}}]{vijay2007}%
  \BibitemOpen
  \bibfield  {author} {\bibinfo {author} {\bibfnamefont {V.}~\bibnamefont
  {Narayan}}, \bibinfo {author} {\bibfnamefont {S.}~\bibnamefont {Ramaswamy}},
  \ and\ \bibinfo {author} {\bibfnamefont {N.}~\bibnamefont {Menon}},\ }\href
  {\doibase 10.1126/science.1140414} {\bibfield  {journal} {\bibinfo  {journal}
  {Science}\ }\textbf {\bibinfo {volume} {317}},\ \bibinfo {pages} {105}
  (\bibinfo {year} {2007})}\BibitemShut {NoStop}%
\bibitem [{\citenamefont {Giavazzi}\ \emph {et~al.}(2017)\citenamefont
  {Giavazzi}, \citenamefont {Malinverno}, \citenamefont {Corallino},
  \citenamefont {Ginelli}, \citenamefont {Scita},\ and\ \citenamefont
  {Cerbino}}]{giavazzi2017}%
  \BibitemOpen
  \bibfield  {author} {\bibinfo {author} {\bibfnamefont {F.}~\bibnamefont
  {Giavazzi}}, \bibinfo {author} {\bibfnamefont {C.}~\bibnamefont
  {Malinverno}}, \bibinfo {author} {\bibfnamefont {S.}~\bibnamefont
  {Corallino}}, \bibinfo {author} {\bibfnamefont {F.}~\bibnamefont {Ginelli}},
  \bibinfo {author} {\bibfnamefont {G.}~\bibnamefont {Scita}}, \ and\ \bibinfo
  {author} {\bibfnamefont {R.}~\bibnamefont {Cerbino}},\ }\href {\doibase
  10.1088/1361-6463/aa7f8e} {\bibfield  {journal} {\bibinfo  {journal} {J.
  Phys. D: Appl. Phys.}\ }\textbf {\bibinfo {volume} {50}},\ \bibinfo {pages}
  {384003} (\bibinfo {year} {2017})}\BibitemShut {NoStop}%
\bibitem [{\citenamefont {Caprini}\ \emph {et~al.}(2020)\citenamefont
  {Caprini}, \citenamefont {Marconi}, \citenamefont {Maggi}, \citenamefont
  {Paoluzzi},\ and\ \citenamefont {Puglisi}}]{caprini2020}%
  \BibitemOpen
  \bibfield  {author} {\bibinfo {author} {\bibfnamefont {L.}~\bibnamefont
  {Caprini}}, \bibinfo {author} {\bibfnamefont {U.~M.~B.}\ \bibnamefont
  {Marconi}}, \bibinfo {author} {\bibfnamefont {C.}~\bibnamefont {Maggi}},
  \bibinfo {author} {\bibfnamefont {M.}~\bibnamefont {Paoluzzi}}, \ and\
  \bibinfo {author} {\bibfnamefont {A.}~\bibnamefont {Puglisi}},\ }\href
  {\doibase 10.1103/PhysRevResearch.2.023321} {\bibfield  {journal} {\bibinfo
  {journal} {Phys. Rev. Res.}\ }\textbf {\bibinfo {volume} {2}},\ \bibinfo
  {pages} {023321} (\bibinfo {year} {2020})}\BibitemShut {NoStop}%
\bibitem [{\citenamefont {Berthier}\ \emph {et~al.}(2019)\citenamefont
  {Berthier}, \citenamefont {Flenner},\ and\ \citenamefont
  {Szamel}}]{berthier2019}%
  \BibitemOpen
  \bibfield  {author} {\bibinfo {author} {\bibfnamefont {L.}~\bibnamefont
  {Berthier}}, \bibinfo {author} {\bibfnamefont {E.}~\bibnamefont {Flenner}}, \
  and\ \bibinfo {author} {\bibfnamefont {G.}~\bibnamefont {Szamel}},\ }\href
  {\doibase 10.1063/1.5093240} {\bibfield  {journal} {\bibinfo  {journal} {J.
  Chem. Phys.}\ }\textbf {\bibinfo {volume} {150}},\ \bibinfo {pages} {200901}
  (\bibinfo {year} {2019})}\BibitemShut {NoStop}%
\bibitem [{\citenamefont {Karmakar}\ \emph {et~al.}(2012)\citenamefont
  {Karmakar}, \citenamefont {Lerner},\ and\ \citenamefont
  {Procaccia}}]{skPhysica2012}%
  \BibitemOpen
  \bibfield  {author} {\bibinfo {author} {\bibfnamefont {S.}~\bibnamefont
  {Karmakar}}, \bibinfo {author} {\bibfnamefont {E.}~\bibnamefont {Lerner}}, \
  and\ \bibinfo {author} {\bibfnamefont {I.}~\bibnamefont {Procaccia}},\ }\href
  {\doibase https://doi.org/10.1016/j.physa.2011.11.020} {\bibfield  {journal}
  {\bibinfo  {journal} {Physica A: Statistical Mechanics and its Applications}\
  }\textbf {\bibinfo {volume} {391}},\ \bibinfo {pages} {1001} (\bibinfo {year}
  {2012})}\BibitemShut {NoStop}%
\bibitem [{\citenamefont {Cammarota}\ and\ \citenamefont
  {Biroli}(2012)}]{chiara2012}%
  \BibitemOpen
  \bibfield  {author} {\bibinfo {author} {\bibfnamefont {C.}~\bibnamefont
  {Cammarota}}\ and\ \bibinfo {author} {\bibfnamefont {G.}~\bibnamefont
  {Biroli}},\ }\href {\doibase 10.1073/pnas.1111582109} {\bibfield  {journal}
  {\bibinfo  {journal} {Proc. Natl. Acad. Sci. (USA)}\ }\textbf {\bibinfo
  {volume} {109}},\ \bibinfo {pages} {8850} (\bibinfo {year}
  {2012})}\BibitemShut {NoStop}%
\bibitem [{\citenamefont {Chakrabarty}\ \emph {et~al.}(2015)\citenamefont
  {Chakrabarty}, \citenamefont {Karmakar},\ and\ \citenamefont
  {Dasgupta}}]{sourishSciReport}%
  \BibitemOpen
  \bibfield  {author} {\bibinfo {author} {\bibfnamefont {S.}~\bibnamefont
  {Chakrabarty}}, \bibinfo {author} {\bibfnamefont {S.}~\bibnamefont
  {Karmakar}}, \ and\ \bibinfo {author} {\bibfnamefont {C.}~\bibnamefont
  {Dasgupta}},\ }\href {\doibase https://doi.org/10.1038/srep12577} {\bibfield
  {journal} {\bibinfo  {journal} {Scientific reports}\ }\textbf {\bibinfo
  {volume} {5}},\ \bibinfo {pages} {12577} (\bibinfo {year}
  {2015})}\BibitemShut {NoStop}%
\bibitem [{\citenamefont {Das}\ \emph {et~al.}(2017)\citenamefont {Das},
  \citenamefont {Chakrabarty},\ and\ \citenamefont
  {Karmakar}}]{softMatter2017}%
  \BibitemOpen
  \bibfield  {author} {\bibinfo {author} {\bibfnamefont {R.}~\bibnamefont
  {Das}}, \bibinfo {author} {\bibfnamefont {S.}~\bibnamefont {Chakrabarty}}, \
  and\ \bibinfo {author} {\bibfnamefont {S.}~\bibnamefont {Karmakar}},\ }\href
  {\doibase 10.1039/C7SM01202K} {\bibfield  {journal} {\bibinfo  {journal}
  {Soft Matter}\ }\textbf {\bibinfo {volume} {13}},\ \bibinfo {pages} {6929}
  (\bibinfo {year} {2017})}\BibitemShut {NoStop}%
\bibitem [{\citenamefont {Ozawa}\ \emph {et~al.}(2015)\citenamefont {Ozawa},
  \citenamefont {Kob}, \citenamefont {Ikeda},\ and\ \citenamefont
  {Miyazaki}}]{OzawaPNAS}%
  \BibitemOpen
  \bibfield  {author} {\bibinfo {author} {\bibfnamefont {M.}~\bibnamefont
  {Ozawa}}, \bibinfo {author} {\bibfnamefont {W.}~\bibnamefont {Kob}}, \bibinfo
  {author} {\bibfnamefont {A.}~\bibnamefont {Ikeda}}, \ and\ \bibinfo {author}
  {\bibfnamefont {K.}~\bibnamefont {Miyazaki}},\ }\href {\doibase
  https://doi.org/10.1073/pnas.1500730112} {\bibfield  {journal} {\bibinfo
  {journal} {Proc. Natl. Acad. Sci. (USA)}\ }\textbf {\bibinfo {volume}
  {112}},\ \bibinfo {pages} {6914} (\bibinfo {year} {2015})}\BibitemShut
  {NoStop}%
\bibitem [{\citenamefont {Prost}\ \emph {et~al.}(2015)\citenamefont {Prost},
  \citenamefont {J{\"{u}}licher},\ and\ \citenamefont {Joanny}}]{jacques2015}%
  \BibitemOpen
  \bibfield  {author} {\bibinfo {author} {\bibfnamefont {J.}~\bibnamefont
  {Prost}}, \bibinfo {author} {\bibfnamefont {F.}~\bibnamefont
  {J{\"{u}}licher}}, \ and\ \bibinfo {author} {\bibfnamefont {J.~F.}\
  \bibnamefont {Joanny}},\ }\href {\doibase 10.1038/nphys3224} {\bibfield
  {journal} {\bibinfo  {journal} {Nat. Phys.}\ }\textbf {\bibinfo {volume}
  {11}},\ \bibinfo {pages} {111} (\bibinfo {year} {2015})}\BibitemShut
  {NoStop}%
\bibitem [{\citenamefont {Karmakar}\ \emph {et~al.}(2009)\citenamefont
  {Karmakar}, \citenamefont {Dasgupta},\ and\ \citenamefont
  {Sastry}}]{smarajitPNAS2009}%
  \BibitemOpen
  \bibfield  {author} {\bibinfo {author} {\bibfnamefont {S.}~\bibnamefont
  {Karmakar}}, \bibinfo {author} {\bibfnamefont {C.}~\bibnamefont {Dasgupta}},
  \ and\ \bibinfo {author} {\bibfnamefont {S.}~\bibnamefont {Sastry}},\ }\href
  {\doibase 10.1073/pnas.0811082106} {\bibfield  {journal} {\bibinfo  {journal}
  {Proc. Natl. Acad. Sci. (USA)}\ }\textbf {\bibinfo {volume} {106}},\ \bibinfo
  {pages} {3675} (\bibinfo {year} {2009})}\BibitemShut {NoStop}%
\bibitem [{\citenamefont {Chakrabarty}\ \emph {et~al.}(2017)\citenamefont
  {Chakrabarty}, \citenamefont {Tah}, \citenamefont {Karmakar},\ and\
  \citenamefont {Dasgupta}}]{saurish2017}%
  \BibitemOpen
  \bibfield  {author} {\bibinfo {author} {\bibfnamefont {S.}~\bibnamefont
  {Chakrabarty}}, \bibinfo {author} {\bibfnamefont {I.}~\bibnamefont {Tah}},
  \bibinfo {author} {\bibfnamefont {S.}~\bibnamefont {Karmakar}}, \ and\
  \bibinfo {author} {\bibfnamefont {C.}~\bibnamefont {Dasgupta}},\ }\href
  {\doibase 10.1103/PhysRevLett.119.205502} {\bibfield  {journal} {\bibinfo
  {journal} {Phys. Rev. Lett.}\ }\textbf {\bibinfo {volume} {119}},\ \bibinfo
  {pages} {205502} (\bibinfo {year} {2017})}\BibitemShut {NoStop}%
\bibitem [{\citenamefont {Tah}\ \emph {et~al.}(2021)\citenamefont {Tah},
  \citenamefont {Mutneja},\ and\ \citenamefont {Karmakar}}]{ACSOmegaReview}%
  \BibitemOpen
  \bibfield  {author} {\bibinfo {author} {\bibfnamefont {I.}~\bibnamefont
  {Tah}}, \bibinfo {author} {\bibfnamefont {A.}~\bibnamefont {Mutneja}}, \ and\
  \bibinfo {author} {\bibfnamefont {S.}~\bibnamefont {Karmakar}},\ }\href
  {\doibase https://doi.org/10.1021/acsomega.0c04831} {\bibfield  {journal}
  {\bibinfo  {journal} {ACS Omega}\ }\textbf {\bibinfo {volume} {6}},\ \bibinfo
  {pages} {7229–7239} (\bibinfo {year} {2021})}\BibitemShut {NoStop}%
\bibitem [{\citenamefont {Kirkpatrick}\ and\ \citenamefont
  {Thirumalai}(2015)}]{kirkpatrick2015}%
  \BibitemOpen
  \bibfield  {author} {\bibinfo {author} {\bibfnamefont {T.~R.}\ \bibnamefont
  {Kirkpatrick}}\ and\ \bibinfo {author} {\bibfnamefont {D.}~\bibnamefont
  {Thirumalai}},\ }\href {\doibase 10.1103/RevModPhys.87.183} {\bibfield
  {journal} {\bibinfo  {journal} {Rev. Mod. Phys.}\ }\textbf {\bibinfo {volume}
  {87}},\ \bibinfo {pages} {183} (\bibinfo {year} {2015})}\BibitemShut
  {NoStop}%
\bibitem [{\citenamefont {Kirkpatrick}\ \emph {et~al.}(1989)\citenamefont
  {Kirkpatrick}, \citenamefont {Thirumalai},\ and\ \citenamefont
  {Wolynes}}]{kirkpatrick1989}%
  \BibitemOpen
  \bibfield  {author} {\bibinfo {author} {\bibfnamefont {T.~R.}\ \bibnamefont
  {Kirkpatrick}}, \bibinfo {author} {\bibfnamefont {D.}~\bibnamefont
  {Thirumalai}}, \ and\ \bibinfo {author} {\bibfnamefont {P.~G.}\ \bibnamefont
  {Wolynes}},\ }\href {\doibase 10.1103/PhysRevA.40.1045} {\bibfield  {journal}
  {\bibinfo  {journal} {Phys. Rev. A}\ }\textbf {\bibinfo {volume} {40}},\
  \bibinfo {pages} {1045} (\bibinfo {year} {1989})}\BibitemShut {NoStop}%
\bibitem [{\citenamefont {Mandal}\ \emph {et~al.}(2020)\citenamefont {Mandal},
  \citenamefont {Bhuyan}, \citenamefont {Chaudhuri}, \citenamefont {Dasgupta},\
  and\ \citenamefont {Rao}}]{mandal2020}%
  \BibitemOpen
  \bibfield  {author} {\bibinfo {author} {\bibfnamefont {R.}~\bibnamefont
  {Mandal}}, \bibinfo {author} {\bibfnamefont {P.~J.}\ \bibnamefont {Bhuyan}},
  \bibinfo {author} {\bibfnamefont {P.}~\bibnamefont {Chaudhuri}}, \bibinfo
  {author} {\bibfnamefont {C.}~\bibnamefont {Dasgupta}}, \ and\ \bibinfo
  {author} {\bibfnamefont {M.}~\bibnamefont {Rao}},\ }\href {\doibase
  10.1038/s41467-020-16130-x} {\bibfield  {journal} {\bibinfo  {journal} {Nat.
  Comm.}\ }\textbf {\bibinfo {volume} {11}},\ \bibinfo {pages} {2581} (\bibinfo
  {year} {2020})}\BibitemShut {NoStop}%
\bibitem [{\citenamefont {Mandal}\ \emph {et~al.}(2021)\citenamefont {Mandal},
  \citenamefont {Nandi}, \citenamefont {Dasgupta}, \citenamefont {Sollich},\
  and\ \citenamefont {Gov}}]{rituparno2021}%
  \BibitemOpen
  \bibfield  {author} {\bibinfo {author} {\bibfnamefont {R.}~\bibnamefont
  {Mandal}}, \bibinfo {author} {\bibfnamefont {S.~K.}\ \bibnamefont {Nandi}},
  \bibinfo {author} {\bibfnamefont {C.}~\bibnamefont {Dasgupta}}, \bibinfo
  {author} {\bibfnamefont {P.}~\bibnamefont {Sollich}}, \ and\ \bibinfo
  {author} {\bibfnamefont {N.~S.}\ \bibnamefont {Gov}},\ }\href
  {https://arxiv.org/abs/2102.07519} {\bibfield  {journal} {\bibinfo  {journal}
  {arXiv: 2102.07519}\ } (\bibinfo {year} {2021})}\BibitemShut {NoStop}%
\bibitem [{\citenamefont {Berthier}\ and\ \citenamefont
  {Witten}(2009)}]{berthier2009}%
  \BibitemOpen
  \bibfield  {author} {\bibinfo {author} {\bibfnamefont {L.}~\bibnamefont
  {Berthier}}\ and\ \bibinfo {author} {\bibfnamefont {T.~A.}\ \bibnamefont
  {Witten}},\ }\href {\doibase 10.1103/PhysRevE.80.021502} {\bibfield
  {journal} {\bibinfo  {journal} {Phys. Rev. E}\ }\textbf {\bibinfo {volume}
  {80}},\ \bibinfo {pages} {021502} (\bibinfo {year} {2009})}\BibitemShut
  {NoStop}%
\bibitem [{\citenamefont {Adhikari}\ \emph {et~al.}(2021)\citenamefont
  {Adhikari}, \citenamefont {Karmakar},\ and\ \citenamefont
  {Sastry}}]{monojJPCB}%
  \BibitemOpen
  \bibfield  {author} {\bibinfo {author} {\bibfnamefont {M.}~\bibnamefont
  {Adhikari}}, \bibinfo {author} {\bibfnamefont {S.}~\bibnamefont {Karmakar}},
  \ and\ \bibinfo {author} {\bibfnamefont {S.}~\bibnamefont {Sastry}},\ }\href
  {\doibase 10.1021/acs.jpcb.1c03887} {\bibfield  {journal} {\bibinfo
  {journal} {The Journal of Physical Chemistry B}\ }\textbf {\bibinfo {volume}
  {125}},\ \bibinfo {pages} {10232} (\bibinfo {year} {2021})},\ \bibinfo {note}
  {pMID: 34494429},\ \Eprint
  {http://arxiv.org/abs/https://doi.org/10.1021/acs.jpcb.1c03887}
  {https://doi.org/10.1021/acs.jpcb.1c03887} \BibitemShut {NoStop}%
\bibitem [{\citenamefont {Sadhukan}\ and\ \citenamefont
  {Nandi}(2021)}]{souvik2021}%
  \BibitemOpen
  \bibfield  {author} {\bibinfo {author} {\bibfnamefont {S.}~\bibnamefont
  {Sadhukan}}\ and\ \bibinfo {author} {\bibfnamefont {S.~K.}\ \bibnamefont
  {Nandi}},\ }\href {\doibase 10.1103/PhysRevE.103.062403} {\bibfield
  {journal} {\bibinfo  {journal} {Phys. Rev. E}\ }\textbf {\bibinfo {volume}
  {103}},\ \bibinfo {pages} {062403} (\bibinfo {year} {2021})}\BibitemShut
  {NoStop}%
\end{thebibliography}%


\begin{thebibliography}{6}%
\makeatletter
\providecommand \@ifxundefined [1]{%
 \@ifx{#1\undefined}
}%
\providecommand \@ifnum [1]{%
 \ifnum #1\expandafter \@firstoftwo
 \else \expandafter \@secondoftwo
 \fi
}%
\providecommand \@ifx [1]{%
 \ifx #1\expandafter \@firstoftwo
 \else \expandafter \@secondoftwo
 \fi
}%
\providecommand \natexlab [1]{#1}%
\providecommand \enquote  [1]{``#1''}%
\providecommand \bibnamefont  [1]{#1}%
\providecommand \bibfnamefont [1]{#1}%
\providecommand \citenamefont [1]{#1}%
\providecommand \href@noop [0]{\@secondoftwo}%
\providecommand \href [0]{\begingroup \@sanitize@url \@href}%
\providecommand \@href[1]{\@@startlink{#1}\@@href}%
\providecommand \@@href[1]{\endgroup#1\@@endlink}%
\providecommand \@sanitize@url [0]{\catcode `\\12\catcode `\$12\catcode
  `\&12\catcode `\#12\catcode `\^12\catcode `\_12\catcode `\%12\relax}%
\providecommand \@@startlink[1]{}%
\providecommand \@@endlink[0]{}%
\providecommand \url  [0]{\begingroup\@sanitize@url \@url }%
\providecommand \@url [1]{\endgroup\@href {#1}{\urlprefix }}%
\providecommand \urlprefix  [0]{URL }%
\providecommand \Eprint [0]{\href }%
\providecommand \doibase [0]{http://dx.doi.org/}%
\providecommand \selectlanguage [0]{\@gobble}%
\providecommand \bibinfo  [0]{\@secondoftwo}%
\providecommand \bibfield  [0]{\@secondoftwo}%
\providecommand \translation [1]{[#1]}%
\providecommand \BibitemOpen [0]{}%
\providecommand \bibitemStop [0]{}%
\providecommand \bibitemNoStop [0]{.\EOS\space}%
\providecommand \EOS [0]{\spacefactor3000\relax}%
\providecommand \BibitemShut  [1]{\csname bibitem#1\endcsname}%
\let\auto@bib@innerbib\@empty
\bibitem [{\citenamefont {Kob}\ and\ \citenamefont {Andersen}(1995)}]{kob1995}%
  \BibitemOpen
  \bibfield  {author} {\bibinfo {author} {\bibfnamefont {W.}~\bibnamefont
  {Kob}}\ and\ \bibinfo {author} {\bibfnamefont {H.~C.}\ \bibnamefont
  {Andersen}},\ }\href {\doibase 10.1103/PhysRevE.51.4626} {\bibfield
  {journal} {\bibinfo  {journal} {Phys. Rev. E}\ }\textbf {\bibinfo {volume}
  {51}},\ \bibinfo {pages} {4626} (\bibinfo {year} {1995})}\BibitemShut
  {NoStop}%
\bibitem [{\citenamefont {Zhang}(1997)}]{zhang1997}%
  \BibitemOpen
  \bibfield  {author} {\bibinfo {author} {\bibfnamefont {F.}~\bibnamefont
  {Zhang}},\ }\href@noop {} {\bibfield  {journal} {\bibinfo  {journal} {J.
  Chem. Phys.}\ }\textbf {\bibinfo {volume} {106}},\ \bibinfo {pages} {6102}
  (\bibinfo {year} {1997})}\BibitemShut {NoStop}%
\bibitem [{\citenamefont {Paul}\ \emph {et~al.}(2021)\citenamefont {Paul},
  \citenamefont {Nandi},\ and\ \citenamefont {Karmakar}}]{kallol2021}%
  \BibitemOpen
  \bibfield  {author} {\bibinfo {author} {\bibfnamefont {K.}~\bibnamefont
  {Paul}}, \bibinfo {author} {\bibfnamefont {S.~K.}\ \bibnamefont {Nandi}}, \
  and\ \bibinfo {author} {\bibfnamefont {S.}~\bibnamefont {Karmakar}},\ }\href
  {https://arxiv.org/abs/2105.12702} {\bibfield  {journal} {\bibinfo  {journal}
  {arXiv: 2105.12702}\ } (\bibinfo {year} {2021})}\BibitemShut {NoStop}%
\bibitem [{\citenamefont {Chakrabarty}\ \emph {et~al.}(2017)\citenamefont
  {Chakrabarty}, \citenamefont {Tah}, \citenamefont {Karmakar},\ and\
  \citenamefont {Dasgupta}}]{saurish2017}%
  \BibitemOpen
  \bibfield  {author} {\bibinfo {author} {\bibfnamefont {S.}~\bibnamefont
  {Chakrabarty}}, \bibinfo {author} {\bibfnamefont {I.}~\bibnamefont {Tah}},
  \bibinfo {author} {\bibfnamefont {S.}~\bibnamefont {Karmakar}}, \ and\
  \bibinfo {author} {\bibfnamefont {C.}~\bibnamefont {Dasgupta}},\ }\href
  {\doibase 10.1103/PhysRevLett.119.205502} {\bibfield  {journal} {\bibinfo
  {journal} {Phys. Rev. Lett.}\ }\textbf {\bibinfo {volume} {119}},\ \bibinfo
  {pages} {205502} (\bibinfo {year} {2017})}\BibitemShut {NoStop}%
\bibitem [{\citenamefont {Karmakar}\ \emph {et~al.}(2009)\citenamefont
  {Karmakar}, \citenamefont {Dasgupta},\ and\ \citenamefont
  {Sastry}}]{smarajitPNAS2009}%
  \BibitemOpen
  \bibfield  {author} {\bibinfo {author} {\bibfnamefont {S.}~\bibnamefont
  {Karmakar}}, \bibinfo {author} {\bibfnamefont {C.}~\bibnamefont {Dasgupta}},
  \ and\ \bibinfo {author} {\bibfnamefont {S.}~\bibnamefont {Sastry}},\ }\href
  {\doibase 10.1073/pnas.0811082106} {\bibfield  {journal} {\bibinfo  {journal}
  {Proc. Natl. Acad. Sci. (USA)}\ }\textbf {\bibinfo {volume} {106}},\ \bibinfo
  {pages} {3675} (\bibinfo {year} {2009})}\BibitemShut {NoStop}%
\bibitem [{\citenamefont {Tah}\ and\ \citenamefont
  {Karmakar}(2021)}]{indrajit2021}%
  \BibitemOpen
  \bibfield  {author} {\bibinfo {author} {\bibfnamefont {I.}~\bibnamefont
  {Tah}}\ and\ \bibinfo {author} {\bibfnamefont {S.}~\bibnamefont {Karmakar}},\
  }\href {https://arxiv.org/pdf/2108.02371.pdf} {\bibfield  {journal} {\bibinfo
   {journal} {arXiv:2108.02371}\ } (\bibinfo {year} {2021})}\BibitemShut
  {NoStop}%
\end{thebibliography}%

\end{document}


\title{Nontrivial activity dependence of static length scale and critical tests of active random first-order transition theory}
	
\author{Kallol Paul}
\affiliation{TIFR Center for Interdisciplinary Science, Tata Institute of Fundamental Research, 36/P Gopanpally Village, Serilingampally Mandal, RR District, Hyderabad, 500075, Telangana, India}
\author{Saroj Kumar Nandi} 
\affiliation{TIFR Center for Interdisciplinary Science, Tata Institute of Fundamental Research, 36/P Gopanpally Village, Serilingampally Mandal, RR District, Hyderabad, 500075, Telangana, India}
\author{Smarajit Karmakar}
\email{smarajit@tifrh.res.in}
\affiliation{TIFR Center for Interdisciplinary Science, Tata Institute of Fundamental Research, 36/P Gopanpally Village, Serilingampally Mandal, RR District, Hyderabad, 500075, Telangana, India}


\maketitle


\section{Model and Methods}
We have studied a well known generic glass-forming liquid  Kob-Andersen binary mixture ({\bf 3dKA model}) in three spacial dimensions. The bi-dispersity was particularly chosen to avoid the crystallization as well as the phase separation and the bi-dispersity ratio is $80 : 20$ ($A : B$) \cite{kob1995}. The two types of particles $\alpha$ and $\beta$ ($\alpha\beta={\rm AA, AB, BB}$) are separated by a distance $r$ from each other and interact via the following Lenard-Jones potential
\begin{equation}
\Phi_{\alpha\beta}(r) = \begin{cases}
4 \varepsilon_{\alpha\beta} \left[\left(\frac{\sigma_{\alpha \beta}}{r}\right)^{12} 
- \left(\frac{\sigma_{\alpha\beta}}{r}\right)^6\right], \quad  \text{if $r \leq r_{\rm c}\sigma_{\alpha\beta}$}\\
0,  \quad\quad \text{if $r \geq r_{\rm c}\sigma_{\alpha\beta}$},
\end{cases}
\end{equation}
where \(r_{\rm c}\) is the interaction cut-off distance chosen at $r_{\rm c}=2.5$ above which the potential vanishes. In this work, energies and lengths are given in units of $\varepsilon_{\rm AA}$ and $\sigma_{\rm AA}$, respectively. The parameters of the Lenard-Jones potential are $\varepsilon_{\rm AB}= 1.5 \varepsilon_{\rm AA}$, $\varepsilon_{\rm BB}=0.5\varepsilon_{\rm AA}$, $\sigma_{\rm AB}=0.8\sigma_{\rm AA}$, and $\sigma_{\rm BB}=0.88 \sigma_{\rm AA}$. The Boltzmann constant, $k_{\rm B}$, as well as the mass of any individual particle, $m$, are set to unity, $k_{\rm B}=1$ and $m=1$. We set the number density of the system at $\rho= N/V =1.2$, where $N$ is the system  size and $V$ is the volume. The number of particles $N$ is varied from $N = 160$ to $N = 50000$ to perform the finite size scaling analysis.

In our simulations, we used the protocol same as the recent study by Paul {\it et al.}~\cite{kallol2021}. We have applied extra force to only  a fraction $\rho_a$ of the total number of particles to introduce activity in the system. These active particles are chosen randomly in the system and assigned a self-propulsion force of the form $\vec{f}= f_0(k_x\hat{x} + k_y\hat{y} + k_z\hat{z})$, where $k_x$, $k_y$, $k_z$ are $\pm1$, chosen randomly so that the net momentum of the system remains zero. After every persistence time $\tau_p$, the direction of the self propulsion force is changed by choosing a different set of values of $k_x$, $k_y$, $k_z$, maintaining the momentum conservation. In this work, we have always used $\rho_a= 0.1$ and $\tau_p= 1.0$. We have only varied the intensity of the active force $f_0$   in the range of $0.0$ to $2.5$ and study the effect of activity as a function of $f_0$ only.

We have performed $NVT$ simulations for different temperatures, activities and system sizes where the equations of motion are integrated via velocity-Verlet integration scheme using an integration time step of $\delta t= 0.005\,\tau$ where $\tau=\sqrt{m \sigma_{\rm AA}^2/\varepsilon_{\rm AA}}$. For every simulation first we have  equilibrated our system at least for $50\tau_{\alpha}$ and then stored the data up to similar simulation time. For $N=50000$, we have performed $10$ statistically independent ensembles for each temperature and activity while for other smaller system sizes we performed $32$ statistically independent ensembles for better averaging. To keep the temperature constant we have used the Gaussian operator-splitting thermostat throughout our simulations \cite{zhang1997}. Note that the usual Berendsen thermostat fails to maintain the temperature at the desired value in the presence of non-equilibrium active forcing, while this Gaussian operator 
splitting thermostat is found to maintain the system at one's desired temperature without any deviation even with non-equilibrium active forcing. 

\section{Methods to obtain Static Length Scale}
We have used two independent methods to compute the static length scale $\xi_S$ for all studied activities. 

\subsection{Block analysis method}

\begin{figure}[!htpb]
\begin{center}
\includegraphics[width=18cm]{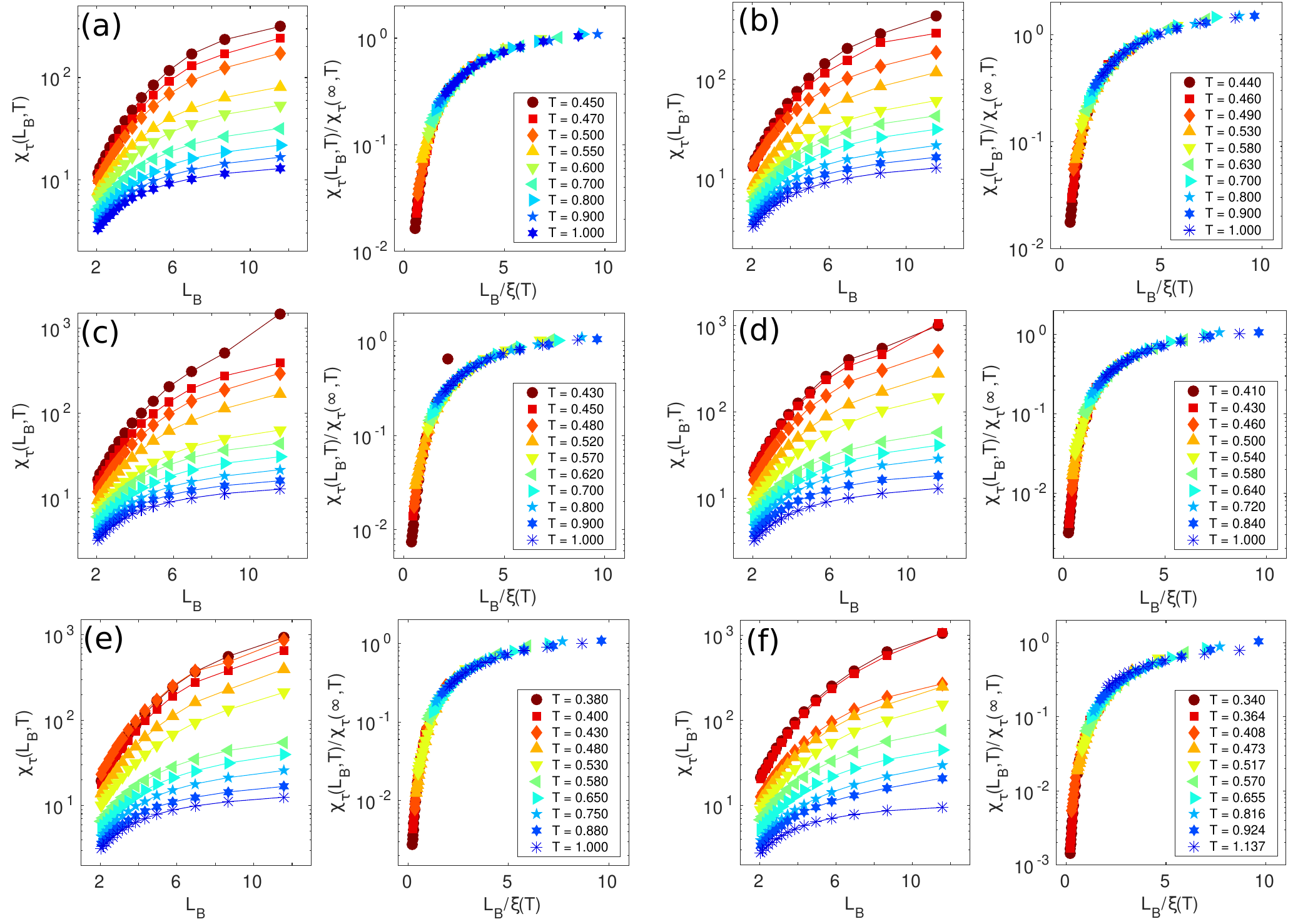}
\caption{Block size dependence of $\chi_\tau$ and finite size scaling for activity $f_0$ = 0.0 (a), $f_0$ = 0.5 (b), $f_0$ = 1.0 (c), $f_0$ = 1.5 (d), $f_0$ = 2.0 (e) and $f_0$ = 2.5 (f)}
\label{chiTau_collapse}
\end{center}
\end{figure}

As one increases the activity at constant temperature, the time scale of the glassy system decreases. To have our observation at a similar time scale for all range of activities, we have chosen a definite temperature window for each activity. We have performed all our MD simulations at that chosen temperature window for corresponding activity. First we used a newly proposed method called ``block analysis'', which is considered an efficient method to perform finite-size scaling with a single sufficiently large system size data for obtaining the static length scale \cite{saurish2017}. In this method, we have considered a small sub-system embedded in a moderately large system of fixed size. We referred these sub-systems as `blocks' and the size of these can vary. Then we looked at the fluctuations of the $\tau_\alpha$ at every different size of these blocks and defined a quantity called $\chi_\tau$ (see definition later). For this method we carried all our simulation at a sufficiently large system size of $N = 50000$ and constructed blocks of size $L_B = L/n$, where $n \in \{3, 4, 5, ...\}$. We computed the $\tau_\alpha$ using the particles which are present inside one such box at a chosen time origin. Then we calculated the distribution of $\tau_\alpha$ as a function of different block size. For each block, we first measured the time at which the overlap correlation function $Q^{(i)}(L_B,t)$ of a particular block for a fixed time origin attains a value of $1/e$. Here the subscript `$i$' refers that the quantity is for a single block $i$ before performing any averaging. From this we further computed $\tau_\alpha^{(i)}(L_B)$ and finally we calculated the mean and variance and defined $\chi_\tau(L_B,T)$ such as,
\begin{equation}
\chi_\tau(L_B,T) = L_B^3\left<\frac{\frac{1}{N_B}\sum_{i=1}^{N_B}[\Delta\tau_\alpha^{(i)}(L_B)]^2}{[\overline{\tau_\alpha^{(i)}(L_B)}]^2}\right>
\end{equation}

where,

\begin{equation}
\overline{\tau_\alpha^{(i)}(L_B)} = \frac{1}{N_B}\sum_{i=1}^{N_B}\tau_\alpha^{(i)}(L_B)
\end{equation}

and

\begin{equation}
\Delta\tau_\alpha^{(i)}(L_B) = \tau_\alpha^{(i)}(L_B) - \overline{\tau_\alpha^{(i)}(L_B)}
\end{equation}
where $N_B$ is the number of blocks with size $L_B$, $n_i$, the number of particles in the $i$th block at time $t = 0$.

\begin{figure}[!htpb]
\begin{center}
\includegraphics[width=18cm]{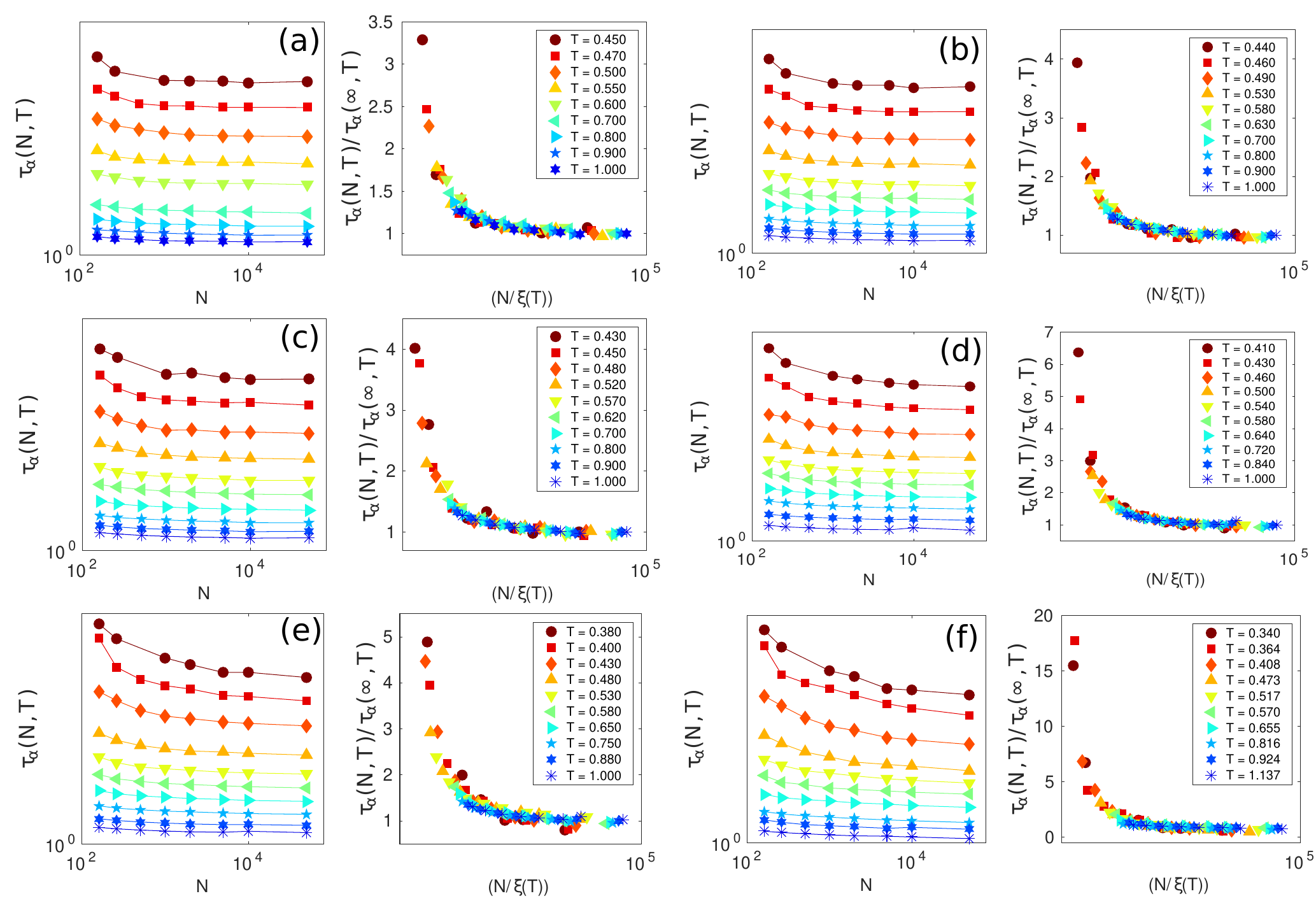}
\caption{System size dependence of $\tau_\alpha$ and finite size scaling for activity $f_0$ = 0.0 (a), 0.5 (b), 1.0 (c), 1.5 (d), 2.0 (e) and 2.5 (f)}
\label{sysSize_collapse}
\end{center}
\end{figure}

For every activity, we considered the variation of $\tau_\alpha$, $\chi_\tau$ at temperature $T$, on the block size  $L_B$ for a fixed value of $N = \rho L^3$. This dependence of $\chi_\tau$ with different block size is shown in Fig.\ref{chiTau_collapse}. The left panel of each sub figure shows the data for $\chi_\tau$ as a function of the block length $L_B$ at different temperature. $\chi_\tau$ at a given $T$ increases with $L_B$ and then saturates at some $T$-dependent value $\chi_\tau(\infty,T)$.
This dependence of $\chi_\tau(L_B,T)$ on $L_B$ is expected to exhibit the following Finite Size Scaling (FSS) form:
\begin{equation}
\chi_\tau(L_B,T) = \chi_0(T)\mathcal{F}\left(\frac{L_B}{\xi_S}\right)
\end{equation}
where,
\begin{equation}
\chi_0(T) = \lim_{L_B\to\infty}\chi_\tau(L_B,T)
\end{equation}
and $\xi_S$ is a characteristic scaling length scale. For each activity the data for all temperatures can be collapsed to a master curve using the two parameters $\chi_\tau(\infty,T)$ and $\xi_S$, as shown in right panel of each sub-figures of Fig. \ref{chiTau_collapse}. The block size dependence of $\chi_\tau$ is governed by the length scale $\xi_S$. This length scale, extracted from this scaling collapse is the same as the static length scale, $\xi_S$, as shown in \cite{smarajitPNAS2009} for the equilibrium glassy system. The excellent data collapse shown in Fig. \ref{chiTau_collapse} confirm that the extracted length scales will be very reliable with small error bars even in the non-equilibrium case.

\subsection{Finite size scaling of $\tau_\alpha$}
We also performed the finite size scaling of $\tau_\alpha$ and extract the static length scale from it. To do this we performed the simulations varying the system size $N = 160$ to $N = 50000$ and computed the corresponding $\tau_\alpha$ for all temperatures, system sizes and activities. This dependence is shown in Fig.\ref{sysSize_collapse}. The left panel of each sub figure shows the data for 
$\tau_\alpha$ as a function of system size $N$ for different temperatures. Fig.\ref{sysSize_collapse} shows that $\tau_\alpha$ at a given temperature decreases with $N$ and saturates at some $T$-dependent value $\tau_\alpha(\infty,T)$. Similarly the dependence of $\tau_\alpha(N,T)$ on $N$ is expected to exhibit the following Finite Size Scaling (FSS) form:
\begin{equation}
\tau_\alpha(N,T) = \tau_0(T)\mathcal{F}\left(\frac{N}{\xi_S}\right)
\end{equation}
where,
\begin{equation}
\tau_0(T) = \lim_{N\to\infty}\tau_\alpha(N,T)
\end{equation}

and $\xi_S$ is a characteristic scaling length scale. The data for all temperatures can be collapsed to a master curve using the two parameters $\tau_\alpha(\infty,T)$ and $\xi_S$, at each temperature, as shown in right panel of each sub-figures of Fig. \ref{sysSize_collapse} for different activities. This clearly show that the associated static length scale is consistently increasing with increasing activity at similar structural relaxation timescale.

\section{Temperature dependence of static length scale}
\begin{figure}[htpb!]
\begin{center}
\includegraphics[scale=0.7]{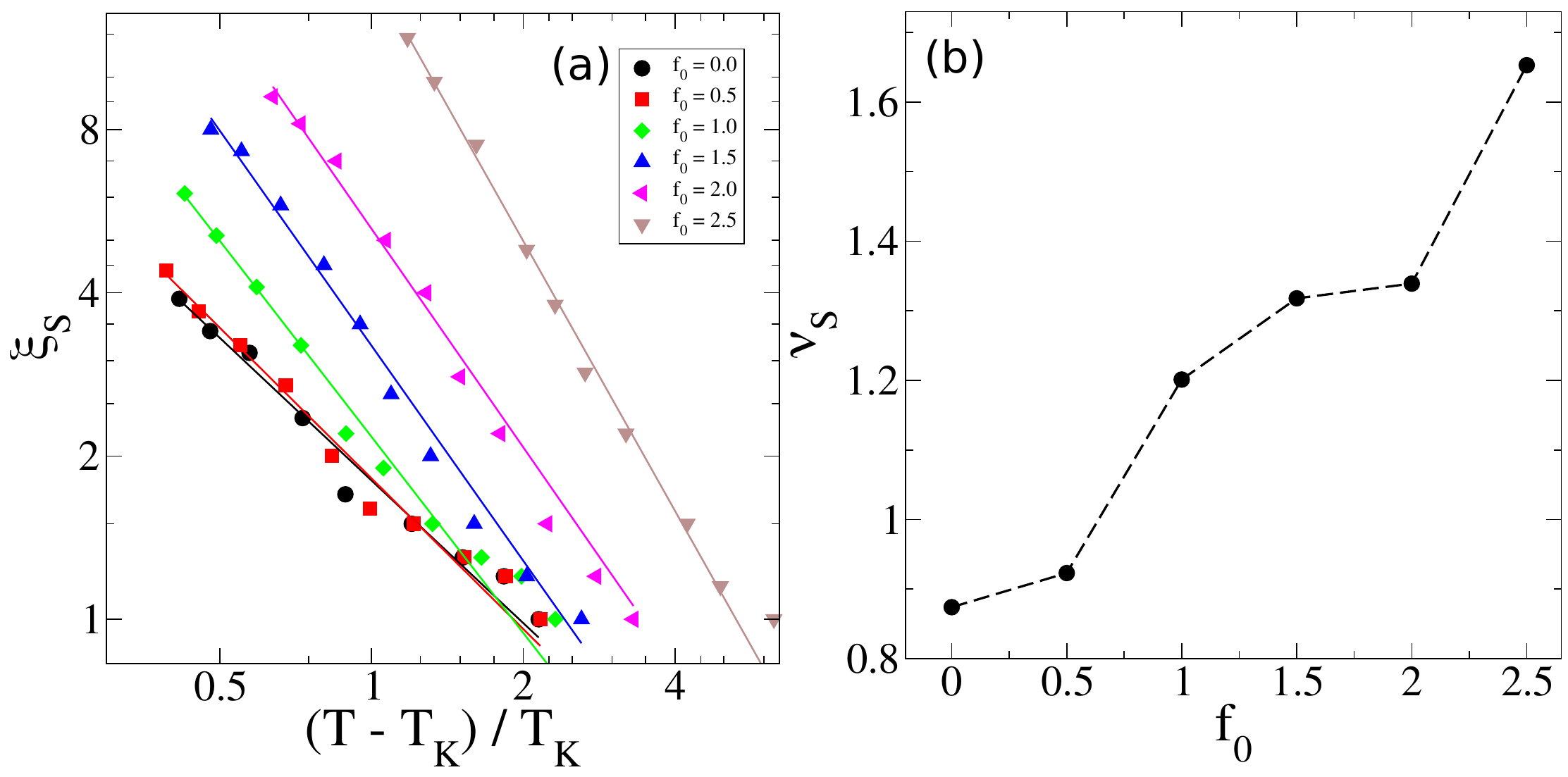}
\caption{(a) Static length scale $\xi_S$ as a function of $(T-T_K)/T_K$; lines are fits with the function $\xi_S\sim |(T-T_K)/T_K|^{-\nu_S}$. (b) The exponent $\nu_S$ consistently increases with $f_0$.}
\label{xaiSvsTbyTk-1}
\end{center}
\end{figure}

Here we have presented the static length scale $\xi_S$ as a function of $|\frac{T-T_K}{T_K}|$ for different activities in Fig. \ref{xaiSvsTbyTk-1}(a) where $T_K$ is the Vogel-Fulcher-Tamman (VFT) divergence temperature, the temperature at which the relaxation time extrapolates to infinity. We obtained the value of $T_K$ by fitting the $\tau_\alpha(T)$ data at different $f_0$ with the VFT equation: $$\tau_\alpha\sim \tau_0\exp{\left(\frac{A}{T-T_K}\right)}.$$ with $\tau_0$, $A$ and $T_K$ being the fitting parameters. Fig. \ref{xaiSvsTbyTk-1}(a) clearly shows that $\xi_S$ seems to show a power-law type divergence with temperature. We computed the exponent $\nu_S$ of this power law for each $f_0$ using the relation $$\xi_S(T) \propto \left|\frac{T-T_K}{T_K}\right|^{-\nu_S}.$$ Fig. \ref{xaiSvsTbyTk-1}(b) shows the exponent $\nu_S$ as a function of $f_0$. The exponent is systematically increasing with increasing $f_0$.

\subsection*{Dependence of $\xi_S$ with $T_g/T$}	
\begin{figure*}[htpb!]
\begin{center}
\includegraphics[scale=0.5]{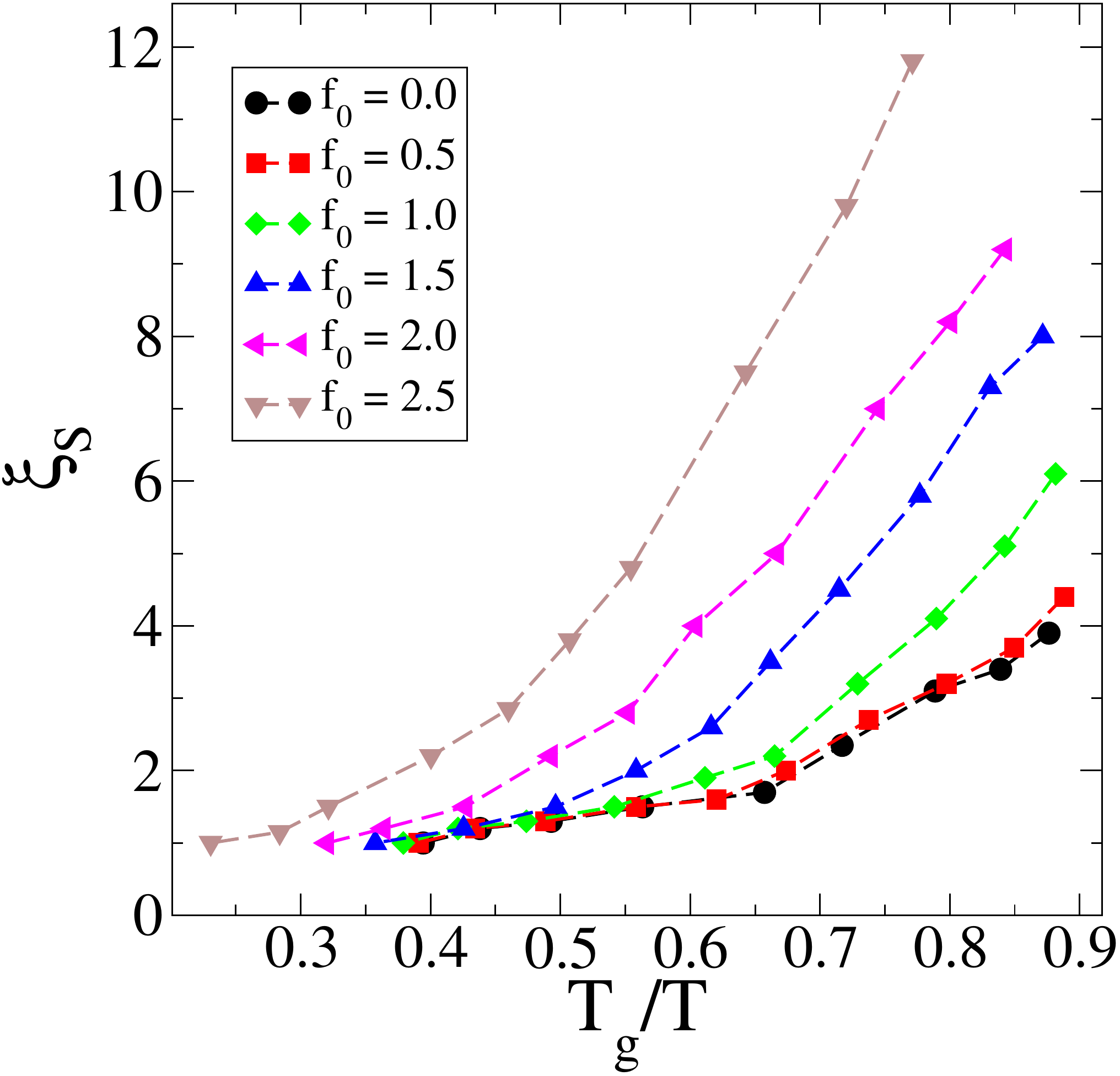}
\caption{Static length scale $\xi_S$ as a function of $T_g/T$ to demonstrate the dramatic growth of length scale within a comparable range of $T_g/T$, where $T_g$ is the calorimetric glass transition temperature, defined as $\tau_\alpha(T_g)=10^6$.}
\label{xaiSvsTgbyT}
\end{center}
\end{figure*}

We have also presented the static length scale, $\xi_S$, as a function of $T_g/T$ in Fig. \ref{xaiSvsTgbyT}, where $T_g$ is the calorimetric glass transition temperature and is defined as $\tau_\alpha(T=T_g)=10^6$. It clearly shows that the static length scale $\xi_S$ is consistently increasing with increasing $f_0$ at similar range of $T_g/T$. Note that with increasing activity the system behaves as a strong liquid even though the static length scale shows stronger temperature dependence. This is very counter intuitive as in equilibrium systems, one finds the temperature dependence of the static length scale to be very mild for strong liquids whereas fragile liquid shows much stronger dependence on temperature \cite{indrajit2021}. This clearly highlights a strong departure from effective equilibrium like behaviour in active glasses when one studies their static structural correlations even if the relaxation time can be described by an effective equilibrium theory. 

\bibliography{activechi4_SIref.bib}